\documentclass[12pt]{article}
\usepackage{epsf}
\usepackage{graphicx}
\usepackage{axodraw}
\usepackage{eepic}
\usepackage{latexsym}
\newcommand{\be}{\begin{equation}}
\newcommand{\ee}{\end{equation}}
\newcommand{\ba}{\begin{array}}
\newcommand{\ea}{\end{array}}
\newcommand{\bqa}{\begin{eqnarray}}
\newcommand{\eqa}{\end{eqnarray}}
\begin{document}
\begin{center}
{\Large\bf The $\kappa$ resonance in $s$ wave $\pi K$ scatterings
}
\\[10mm]
{\sc H.~Q.~Zheng, Z.~Y.~Zhou,  G.~Y.~Qin\footnote{Address after
Sep.~1st, 2003: Department of Physics, McGill University, 3600
University Street, Montreal, Quebec, H3A 2T8, Canada. },
Z.~G.~Xiao, J.~J.~Wang}
\\[2mm]
{\it  Department of Physics, Peking University, Beijing 100871,
P.~R.~China }
\\[2mm] and \\[2mm]
{\sc N.~Wu}
\\[2mm]
{\it  Institute for High Energy Physics, Chinese Academy of
Science, Beijing 100039, P.~R.~China }
\\[5mm]
\today
\begin{abstract}
A new unitarization approach incorporated with chiral symmetry is
established and applied to study the $\pi K$ elastic scatterings.
 We demonstrate that the $\kappa$
resonance exists, if the scattering length parameter in the I=1/2,
J=0 channel does not deviate much from its value predicted by
chiral perturbation theory. The mass and width of the $\kappa$
resonance is found to be $M_\kappa=594\pm 79MeV$,
$\Gamma_\kappa=724\pm 332MeV$, obtained by fitting the LASS data
up to 1430MeV. Better determination to the pole parameters is
possible if the chiral predictions on scattering lengths are taken
into account.
\end{abstract}
\end{center}
\vspace{1cm}
\section {Introduction}\label{intro}

There have been many works devoted to study the resonance
structure in the $I=1/2, J=0$ channel of $\pi K$ scatterings. It
has been suggested  a long time ago that there may exist a
resonance named $\kappa$ in this channel~\cite{Jaffe}. Very
recently, both the E791
 Collaboration~\cite{E791} and the BES Collaboration~\cite{BES}
  have found  evidences
 for  the $\kappa$ resonance, in the $D\to K\pi\pi$ and $J/\Psi\to K^*K\pi$
 channels, respectively, which makes the topic
 even more interesting than ever. Previous
  theoretical studies are mainly based on the LASS data~\cite{ASton} and the
 SLAC data~\cite{Estabrooks} for $\pi K$ scatterings, from which
 the
 $\kappa$  resonance
 was suggested to exist in the $I=1/2, J=0$ channel, based
upon
 various model dependent analysis
 (see for example Refs.~\cite{Beveren86}-- \cite{JOP00}),
 whereas contradicting opinions also exist
 (see for example Refs.~\cite{Torn95}--\cite{CP00}), casting doubt on
 the existence of this resonance.

The reason that different and sometimes contradicting conclusions
exist is partly, if not mainly, due to the fact that, this
channel, like the $I=J=0$ channel in $\pi\pi$ scatterings, is of
strong interaction nature and undergos strong unitarity
corrections. Indeed, $\pi K$ scattering amplitudes have been
calculated up to 1-loop order  in chiral perturbation theory
($\chi$PT)~\cite{Meissner,Vernard}. However, since chiral loop
expansion is an expansion in terms of external momenta and masses
of pseudo-scalar mesons, the perturbation series is in principle
useful only at very low energies due to the violation of unitarity
in perturbation calculation.

This paper is  devoted to the study on  the $\kappa$ resonance in
the $s$ wave $\pi K$ scattering processes.\footnote{Part of the
results are already presented in Ref.~\cite{zhengtalk032}.} For
this purpose, we first develop some new dispersion techniques
which improves and extends our previous results in this
direction~\cite{XZ00,HXZ02,Ztalk03}. The main improvement is that
in our present scheme, unitarity is manifestly preserved.
Different contributions from poles and cuts to the scattering
phase shift  are classified, and different contributions to the
phase shift are additive. Then we make use of the $\chi$PT results
to estimate the left hand cut in the un-physical region. In this
region -- since it is further away from those resonance poles --
$\chi $PT is expected to work well, at least qualitatively. We
find that the $\kappa$ resonance exist, if the scattering length
parameter in the I=1/2 channel does not deviate much from its
value predicted by chiral perturbation theory.

This paper is organized as the following: sec.~\ref{intro} is the
introduction. In sec.~\ref{P}, we review some background knowledge
being used in the later discussions. They include an introduction
to the kinematics for $\pi K$ scatterings and $\chi$PT results at
1--loop level. Especially we discuss the dispersion techniques
previously developed for studying $\pi\pi$
interactions~\cite{XZ00,HXZ02} and modify those dispersion
relations to meet the new kinematics. In sec.~\ref{pade} we make a
pedagogical analysis to the Pad\'e approximation  to the $\chi$PT
amplitudes. This method, and its variations~\cite{Pade}, have been
widely used in the literature to study the non-perturbative
dynamics involving
 pseudo-Goldstone bosons.  However, we reveal  severe problems this
 approximation method encounters in $\pi K$ scatterings,
 similar to what happens in $\pi\pi$ scatterings~\cite{Qin}.
 We
 conclude that in
 dealing with chiral amplitudes, in some cases the Pad\'e approximation is a
 poor unitarization method.
Following the idea proposed in Ref.~\cite{Ztalk03} in
sec.~\ref{NAoU} we develop a new method of unitarization which
respects all known fundamental properties of $S$ matrix theory,
which are unitarity, analyticity and causality, though crossing
symmetry is not implemented automatically. Also efforts have been
made to combine chiral symmetry and the results from chiral
perturbation theory in the new unitarization scheme. The new
unitarization scheme starts from first principles and is formally
rigorous. Of course, approximations have to be made once it is
used in practice, but our scheme clearly shows where those
systematic errors induced by approximations come from -- a
property many models do not maintain. In this section we also
discuss one issue with respect to the Breit--Wigner description of
resonances. Sec.~\ref{fit} devotes to the numerical fit to the
$\pi K$ scattering data. A major topic in this section is the
estimation to the background contributions, which is essential,
and even vital, in studying broad resonances. In the rest of
sec.~\ref{fit} we present detailed numerical analysis on the
location of the $\kappa$ pole with or with out further constraints
from chiral perturbation theory. Sec.~\ref{conclude} is for
discussions and conclusions, including a comparison to several
other related works found in the literature.

\section{Preliminaries}\label{P}
In this section we briefly review some basic properties of $\pi K$
scattering amplitudes. In Sec.~\ref{kinematics}, we introduce the
kinematics for $\pi K$ scatterings. In Sec.~\ref{chiPT} we briefly
review the known results on $\pi K$ scatterings from 1--loop
$SU(3)$ chiral perturbation theory. In Sec.~\ref{DIS} we establish
the dispersion relations for $\pi K$ scatterings following the
method proposed in Ref.~\cite{XZ00}, some of the contents are new.
\subsection {Kinematics for $\pi K$ Scatterings}\label{kinematics}
The center of mass momentum
in the s-channel is written as,%
\be%
k(s) ={1\over {2\sqrt s}}\sqrt{(s-s_R)(s-s_L)}\ ,
\ee%
where
\bqa%
s_R=(m_K+m_{\pi})^2 \ , s_L=(m_K-m_{\pi})^2\ ,
\eqa%
and $\rho(s)$ is the kinematic factor:%
\be%
\rho(s)={{2k(s)}\over {\sqrt s}}={1\over s}\sqrt{(s-s_R)(s-s_L)}\ .%
\ee%
The partial wave $S$ matrix is, %
\be\label{S matrix}%
S(s) = 1+2i\rho(s)T(s)\ , %
\ee%
where and hereafter we  omit the indices of isospin and spin, $I$
and $J$, of the partial wave amplitudes  unless it causes
confusion. In the single channel physical region the $S$ matrix is
unitary which leads to,
\be\label{unitarity}%
\mathrm{Im}{T(s)} = \rho(s) |T(s)|^2\ ,\ (S^\dag S=1)\ .%
\ee%
Since $S$ is unitary in the single channel physical region, the
partial wave $S$ matrix can be conveniently parameterized as the
following,
\be%
S(s) = e^{2i\delta_{1}(s)}\ ,%
\ee%
where $\delta_1$ is the $\pi K$ scattering phase shift.
The subscript $1$ indicates that
$\delta_1$ is only defined, at this moment, in the single channel
physical region, i.e., above the $\pi K$ threshold but below the
$K \eta$ threshold where $\delta_1$ is real. Since
 in the  $I={1\over 2}$ channel the partial wave $S$ matrix
above the $K\eta$ threshold is no longer unitary,  the  $S$ matrix
is then parameterized as the following familiar form,
\bqa%
S(s) = \eta(s)e^{2i\delta_2(s)}\ ,
\eqa%
where $\delta_2$ is the phase shift experimentally measured above
the $K\eta$ threshold, and $\eta(s)$ is the inelasticity
parameter. Apparently $\delta_1$ and $\delta_2$ are different
analytic functions. In fact they have a simple relation above the
second threshold,
 \be\label{d12}
\delta_1(s)=\delta_2(s)-\frac{i}{2}\log\eta(s)\ , \,\,\,{\mathrm
or}\,\,\, \delta_2(s)= {\mathrm Re} \delta_1(s)\,\,\,
(s>(M_K+M_\eta)^2)\ .
 \ee
 The relation Eq.~(\ref{d12}) will be  useful in later discussions. For
 convenience we in the following will also often  use a simpler notation $\delta$ to
 describe the experimentally observed phase shift regardless which
 region it is defined, as is usually adopted in the literature.

\begin{figure}
\setlength{\unitlength}{1.mm}
\begin{center}%
\begin{picture}(140,70)%
\put(0,35){\vector(1,0){140}}%
\put(0,34.8){\vector(1,0){140}}%
\put(0,35.2){\vector(1,0){140}}%

\put(70,0){\vector(0,1){70}}%
\put(69.8,0){\vector(0,1){70}}%
\put(70.2,0){\vector(0,1){70}}%

\put(70,35){\circle{50}}%
\put(70,35){\circle{49.8}}%
\put(70,35){\circle{50.2}}%

\put(70,35){\arc{54}{3.22}{3.06}}%
\put(70,35){\arc{46}{3.22}{3.06}}%

\put(0,33){\line(1,0){43}}%
\put(0,37){\line(1,0){43}}%
\put(47,33){\line(1,0){33}}%
\put(47,37){\line(1,0){33}}%
\put(105,33){\line(1,0){35}}%
\put(105,37){\line(1,0){35}}%

\put(80,35){\oval(4,4)[r]}%
\put(105,35){\oval(4,4)[l]}%

\put(27,33){\vector(-1,0){3.5}}%
\put(23,37){\vector(1,0){3.5}}%
\put(62,33){\vector(-1,0){3.5}}%
\put(58,37){\vector(1,0){3.5}}%
\put(122,33){\vector(-1,0){3.5}}%
\put(118,37){\vector(1,0){3.5}}%

\put(69,12){\vector(1,0){3.5}}%
\put(71,8){\vector(-1,0){3.5}}%
\put(69,62){\vector(1,0){3.5}}%
\put(71,58){\vector(-1,0){3.5}}%

\put(73,25){\scriptsize{$(m_K-m_{\pi})^2$}}
\put(82,28){\vector(0,1){5}}%
\put(89,10){\scriptsize{$m_K^2-m_{\pi}^2$}}
\put(95,15){\vector(0,1){15}}%
\put(96,25){\scriptsize{$(m_K+m_{\pi})^2$}}
\put(103,28){\vector(0,1){5}}%
\end{picture}%
\end{center}%
\caption{\label{figSS} The left hand cut, circular cut and the
right hand cut of $\pi K$ scatterings.}
\end{figure}
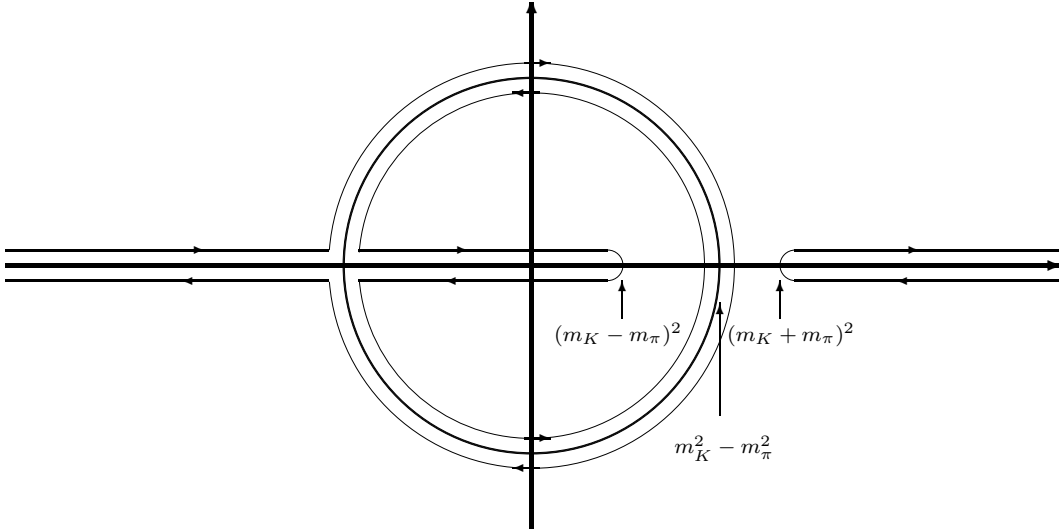

Because $\pi K$ scattering is an unequal mass scattering process,
the singularity structure of its partial wave amplitude is more
complicated than the equal mass scattering.
 The cut structure for the partial wave amplitudes  is
depicted in fig.~\ref{figSS}.~\cite{KS}

\subsection {$\pi K$ scattering amplitudes in SU(3) chiral
perturbation theory}\label{chiPT}

Since the $\pi$ and $K$ has isospin $1$ and $1\over 2$,
respectively, there are two independent $\pi K$ scattering
amplitudes: $T_{\pi K}^{3\over 2}$ and $T_{\pi K}^{1\over 2}$. To
be specific, consider the process which
has purely isospin $3\over 2$,%
\bqa\label{I3 piK scattering process}%
\pi^+(p_1)+K^+(p_2)\rightarrow \pi^+(p_3)+K^+(p_4)\ .%
\eqa%
The amplitude $T_{\pi K}^{3\over 2}$ depends on the conventional
Mandelstam variables
\bqa%
s=(p_1+p_2)^2\ , \ t=(p_1-p_3)^2\ , \ u=(p_1-p_4)^2\ , \eqa
 with
the constraint $s+t+u=2(m_{\pi}^2+m_K^2)$. The
$s\longleftrightarrow u$ crossing of the amplitude Eq.~(\ref{I3
piK scattering process}) will generate the process $\pi^+ K^-
\rightarrow \pi^+ K^-$, which has both $I={3\over 2}$ and
$I={1\over 2}$ components, so we have the following relation:%
\bqa\label{isospin amplitudes relation}%
T_{\pi K}^{1\over 2}(s,t,u) &=& {3\over 2}T_{\pi K}^{3\over
2}(u,t,s)-{1\over 2}T_{\pi K}^{3\over 2}(s,t,u)\ .%
\eqa%
The amplitude $T_{\pi K}^{3\over 2}$ in $SU(3)\times SU(3)$ chiral
perturbation theory was given in Refs.~\cite{Meissner,Vernard},
\bqa\label{T-piK I3}%
T_{\pi K}^{3\over
2}(s,t,u)&=T_2(s,t,u)+T_4^T(s,t,u)+T_4^P(s,t,u)+T_4^U(s,t,u)\ ,%
\eqa%
where $T_2$ denotes the tree-level part, $T_4^T$ denotes the
tadpole terms of order $E^4$, $T_4^P$ denotes the polynomial terms
of order $E^4$ and $T_4^U$ denotes the unitary corrections:%
\bqa\label{T2 T4T T4P T4U}%
T_2(s,t,u) &=& {1\over {2F_{\pi}F_K}}(m_{\pi}^2+m_K^2-s), \nonumber \\%
T_4^T(s,t,u) &=& {1\over
{16F_{\pi}F_K}}(m_K^2-m_{\pi}^2)(-3\mu_{\pi}+2\mu_K+\mu_{\eta}) \nonumber \\%
T_4^P(s,t,u) &=& {2\over
{F_{\pi}^2F_K^2}}(4L_r^1(t-2m_{\pi}^2)(t-2m_K^2)\nonumber \\%
&&+2L_r^2((s-m_{\pi}^2-m_K^2)^2 +(u-m_{\pi}^2-m_K^2)^2)\nonumber \\%
&&+L_r^3((u-m_{\pi}^2-m_K^2)^2+(t-2m_{\pi}^2)(t-2m_K^2))\nonumber \\%
&&+4L_r^4(t(m_{\pi}^2+m_K^2)-4m_{\pi}^2m_K^2)\nonumber \\%
&&+2L_r^5m_{\pi}^2(m_{\pi}^2-m_K^2-s)+8(2L_r^6+L_r^8)m_{\pi}^2m_K^2)\nonumber \\%
T_4^U(s,t,u) &=& {1\over
{4F_{\pi}^2F_K^2}}(t(u-s)(2M_{\pi\pi}^r(t)+M_{KK}^r(t))\nonumber \\%
&&+{3\over 2}((s-t)(L_{\pi K}(u)+L_{K\eta}(u)-u(M_{\pi K}^r(u)+M_{K\eta}^r(u)))\nonumber \\%
&&+(m_K^2-m_{\pi}^2)^2(M_{\pi K}^r(u)+M_{K\eta}^r(u)))\nonumber \\%
&&+{1\over 2}(m_K^2-m_{\pi}^2)(K_{\pi K}(u)(5u-2m_{\pi}^2-2m_K^2)\nonumber \\%
&&+K_{K\eta}(u)(3u-2m_{\pi^2}-2m_K^2))+J_{\pi K}^r(s)(s-m_{\pi}^2-m_K^2)^2\nonumber \\%
&&+{1\over 8}J_{\pi K}^r(u)(11u^2-12u(m_{\pi}^2+m_K^2)+4(m_{\pi}^2+m_K^2)^2)\nonumber \\%
&&+{3\over 8}J_{K\eta}^r(u)(u-{2\over 3}(m_{\pi}^2+m_K^2))^2+{1\over 2}J_{\pi\pi}^r(t)t(2t-m_{\pi}^2)\nonumber \\%
&&+{3\over 4}J_{KK}^r(t)t^2+{1\over 2}J_{\eta\eta}^r(t)m_{\pi}^2(t-{8\over 9}m_K^2))%
\eqa%
where the explicit expressions of functions $\mu_P$ and
$M_{PQ}^r$, $L_{PQ}$, $K_{PQ}$, $J_{PQ}^r$ are displayed in
Ref.~\cite{Gasser2}. In the above Eq.~(\ref{T2 T4T T4P T4U}) of
isospin $I={3\over 2}$ scattering amplitude, we rewrite the
expressions of $T_2(s,t,u)$ and $T_4^T(s,t,u)$ in terms of
${1\over {F_{\pi}F_K}}$ rather than $1\over F_\pi^2$, following
the conventional wisdom. Besides, two typos of the expression of
$T_4^U(s,t,u)$ in Ref.~\cite{Vernard} are corrected as being done
in Ref.~\cite{JOP00}.

 The partial wave expansion of the isospin
amplitudes is written as%
\be%
T^I(s,t) = 16\pi \sum_J(2J+1)P_J(\cos\theta){T^I_J}(s)\ .%
\ee%
The partial wave amplitudes in $\chi $PT  expanded to $O(p^4)$ are%
\be%
T^I_J(s) = T^I_{J, 2}(s)+T^I_{J,4}(s)\ .%
\ee%
The expressions of the partial wave amplitudes are very
voluminous, and we only give the tree level results of $T^{3\over
2}_{0, 2}(s)$ and $T^{1\over 2}_{0,
2}(s)$, %
\bqa\label{treeEq}
 T^{3\over 2}_{0, 2}(s) &=& \frac{{{m_K}}^2 +
{{m_{\pi}}}^2 - s}{32{F_K}{F_{\pi}}\pi }\ ,\nonumber \\%
T^{1\over 2}_{0, 2}(s) &=& \frac{-3{\left( {{m_K}}^2 -
{{m_{\pi}}}^2 \right) }^2 -
    2\left( {{m_K}}^2 + {{m_{\pi}}}^2 \right) s + 5s^2}{128{F_K}{F_{\pi}}\pi
    s}\ .%
\eqa%
The above equations indicate a partial wave $T$ matrix zero at
$s=m_K^2+m_\pi^2$ in the I=3/2 channel and $s\simeq
m_K^2-m_\pi^2/2$ in the I=1/2 channel.\footnote{The second
equation of Eq.~(\ref{treeEq}) affords another tree level zero on
the negative real axis. However after performing the partial wave
integration, the amplitude receives an imaginary contribution at
1--loop level due to the left hand cut and the $T$ matrix zero
disappears. } At 1--loop level the locations of these Adler zeros
will receive corrections depending on the $L_i$ parameters. These
corrections are however rather small since the chiral expansion
works rather well in the energy region.
\subsection{The Dispersion Representations for $\pi K$ Scattering
Amplitudes}\label{DIS}
Following the method of Refs.~\cite{XZ00,HXZ02}, we define two
functions $\tilde{F}$ and $F$,%
\bqa%
\tilde{F}(s) &=& {1\over 2}{\left(S(s)+{1\over S(s)}\right)}\nonumber \\%
F(s) &=& {1\over {2i\rho(s)}}{\left(S(s)-{1\over S(s)}\right)}%
\eqa%
For isospin $3\over 2$, $\tilde{F}$ and $F$ have no right hand cut
in the energy region we are concerning. For isospin $1\over 2$,
$\tilde{F}$ and $F$ have the cut starting from $(M_K+M_{\eta})^2$,
but the right hand cut is very weak until the $(M_K+M_{\eta'})^2$
threshold is reached. The functions $\tilde{F}$, $F$ define the
analytic continuation  of the phase shift $\delta$ in the
following way:
\bqa%
\mathrm{cos}(2\delta_1(s)) &=& {\tilde F(s)}\ ,\nonumber \\%
\mathrm{sin}(2\delta_1(s)) &=& {\rho(s)F(s)}\ .%
\eqa%
 To evaluate  various contributions to phase shifts from
different singularities, we can construct dispersion relations for
$\tilde F(s)$ and $F(s)$ like what have been done in
Ref.~\cite{XZ00,HXZ02}. However the dispersion relations for $\pi
K$ amplitudes are more complicated than those for $\pi\pi$
scatterings, since in here we have the circular cut at
$|s|=M_K^2-M_\pi^2$ as depicted in fig.~\ref{figSS}. Considering
only $F$ for simplicity, we first take the contour of the left
hand integral outside the circular cut,
\bqa\label{dispersion-relations}%
F(s) &=& F(s_0)+\sum_{s_p,\ out}
{{{-(s-s_0)}\mathrm{Res}[F(s_p)]}\over
{(s-s_p)(s_0-s_p)}}+{{(s-s_0)\over \pi}
{\int_{-\infty}^{-(m_K^2-m_{\pi}^2)}}}{{\mathrm{Im_L}F(z)\over{(z-s)(z-s_0)}}dz}\nonumber \\%
&&+ \frac{(s-{s_0})}{2\pi}\int _{\pi }^{-\pi }
\frac{F(({m_K^2}-{m_{\pi }^2}+\epsilon) {e^{i \theta
}})({m_K^2}-{m_{\pi }^2}){e^{i \theta }}{\mathrm{d}\theta
}}{(({m_K^2}-{m_{\pi}^2})
{e^{i\theta}}-s)(({m_K^2}-{m_{\pi}^2}){e^{i \theta }}-{s_0})}\nonumber \\%
&&+ \frac{(s-{s_0})}{\pi}\int _{{({m_{\eta }}+{m_K})^2}}^{\infty }
\frac{\mathrm{Im_R}F(z)}{(z-s)(z-{s_0})}\mathrm{d} z\ .
\eqa%
In the above relation, the integration on the circular cut is
along the outer edge of the circle $|s|=m_K^2-m_\pi^2$ and the sum
runs over all poles, denoted as $s_p$, outside the circle. If we
consider the whole complex $s$ plane, the contribution from the
inner circular cut will be counteracted by the poles inside the
circular cut. So we have the following relation:
\bqa\label{dispersion-relations2}%
0 &=& \sum_{s_p,\ in} {{{-(s-s_0)}\mathrm{Res}[F(s_p)]}\over
{(s-s_p)(s_0-s_p)}}+{{(s-s_0)\over \pi}
{\int_{-(m_K^2-m_{\pi}^2)}^{(m_K-m_{\pi})^2}}}{{\mathrm{Im_L}F(z)\over{(z-s)(z-s_0)}}dz}\nonumber \\%
&&+ \frac{(s-{s_0})}{2\pi}\int _{-\pi }^{\pi }
\frac{F(({m_K^2}-{m_{\pi }^2}-\epsilon) {e^{i \theta
}})({m_K^2}-{m_{\pi }^2}){e^{i \theta }}{\mathrm{d}\theta
}}{(({m_K^2}-{m_{\pi}^2})
{e^{i\theta}}-s)(({m_K^2}-{m_{\pi}^2}){e^{i \theta }}-{s_0})}\ ,%
\eqa%
where the sum runs over all poles inside the circle. Combining
Eqs.~(\ref{dispersion-relations}) and
(\ref{dispersion-relations2}) we get,
\bqa\label{dispersion-relations3}%
F(s) &=& F(s_0)+\sum_{s_p,\ all}
{{{-(s-s_0)}\mathrm{Res}[F(s_p)]}\over
{(s-s_p)(s_0-s_p)}}+{{(s-s_0)\over \pi}
{\int_{-\infty}^{(m_K-m_{\pi})^2}}}{{\mathrm{Im_L}F(z)\over{(z-s)(z-s_0)}}dz}\nonumber \\%
&&+ \frac{(s-{s_0})}{2\pi i}\int _{\pi }^{-\pi }
\frac{\mathrm{disc} F(z)\mathrm{d}z }{(z-s)(z-{s_0})}+
\frac{(s-{s_0})}{\pi}\int _{{({m_{\eta }}+{m_K})^2}}^{\infty }
\frac{\mathrm{Im_R}F(z)}{(z-s)(z-{s_0})}\mathrm{d} z\ .\nonumber\\
\eqa%
Corresponding relations for $\tilde F$ can be similarly written
down.

It is already possible to use Eq.~(\ref{dispersion-relations3}) to
study the $\kappa$ pole problem, parallel to what is done in
Ref.~\cite{XZ00} to study the $\sigma$ resonance. However in the
following we will adopt another improved method which
automatically incorporate the unitarity constraint.

\section{The Unitarization of the Scattering Amplitudes -- Pad\'e
Approximation}\label{pade}

\begin{figure}%
\begin{center}%
\mbox{\epsfxsize=70mm\epsffile{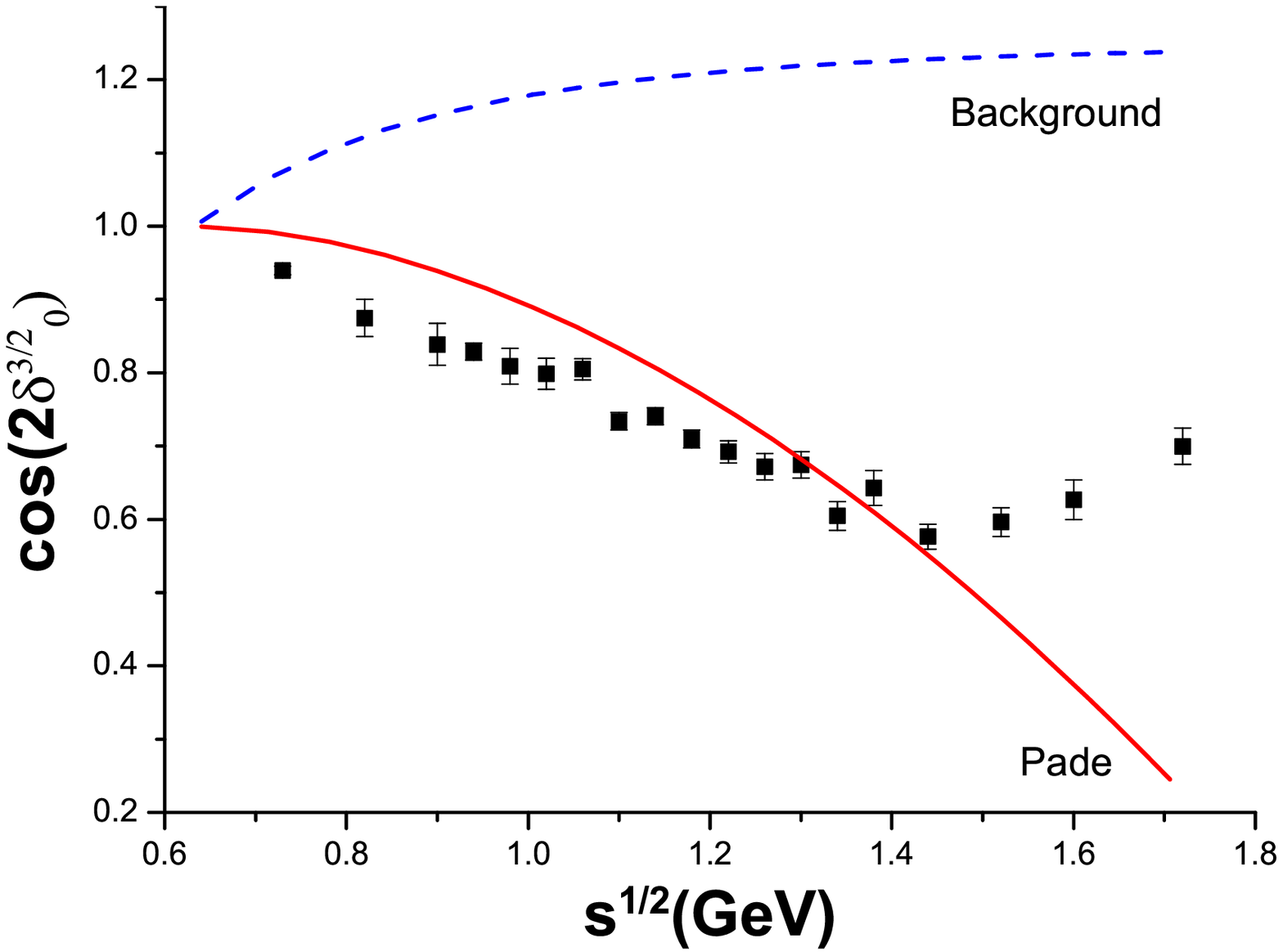}}%
\mbox{\epsfxsize=70mm\epsffile{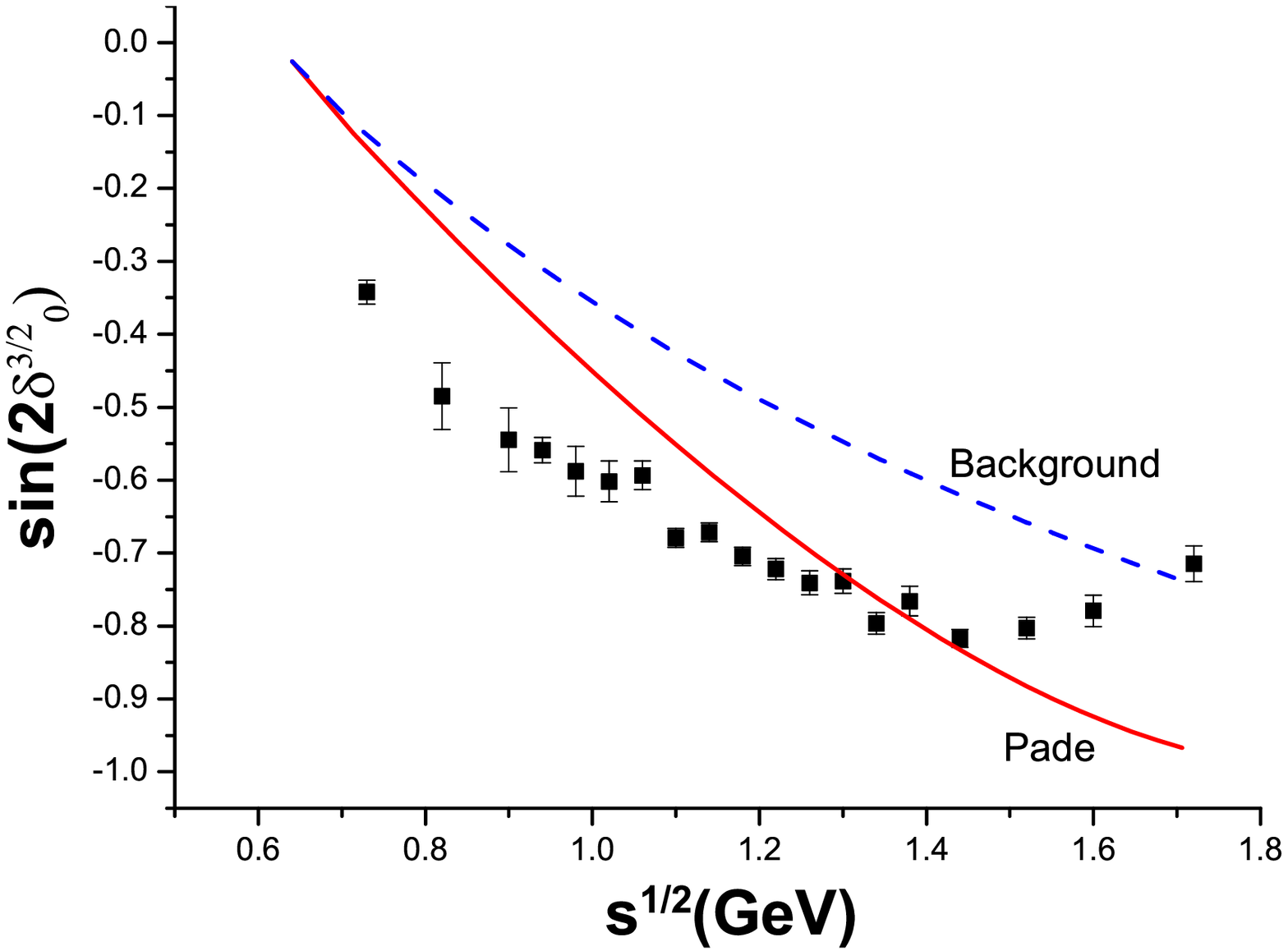}}%
\caption{\label{isospin-3} Various contributions to
$\cos2\delta^{3/2}_0$ and $\sin2\delta^{3/2}_0$ from spurious
poles and left hand cut in the isospin $3\over 2$ $s$ wave Pad\'e
amplitudes. The $L_i$ parameters are taken from the central value
given in  Eq.~(\ref{low-energy constants}). Here `background'
means all contributions on the $r.h.s.$
of Eq.~(\ref{dispersion-relations3}) except the pole terms.}%
\end{center}%
\end{figure}

\begin{figure}%
\begin{center}%
\mbox{\epsfxsize=70mm\epsffile{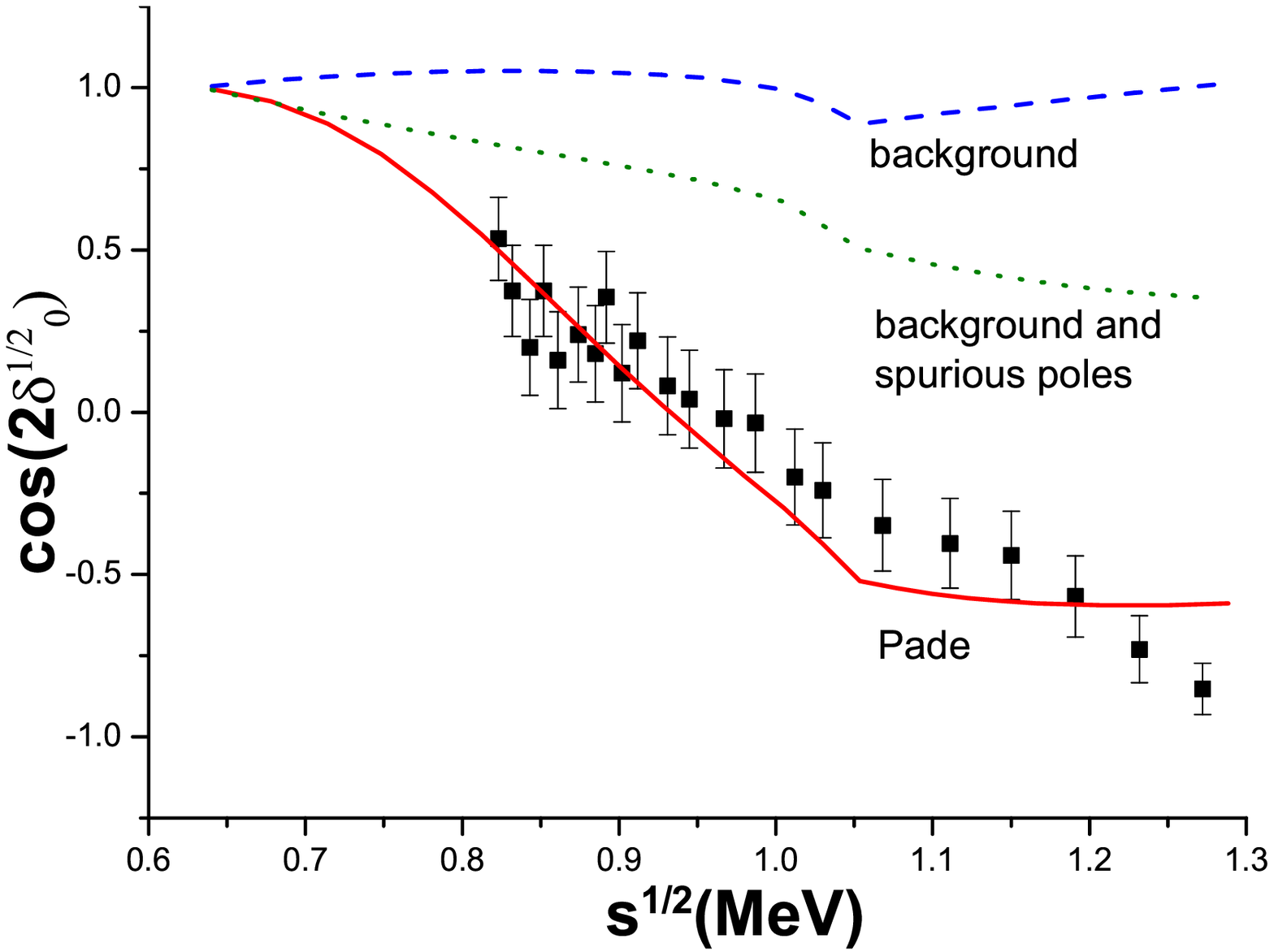}}%
\mbox{\epsfxsize=70mm\epsffile{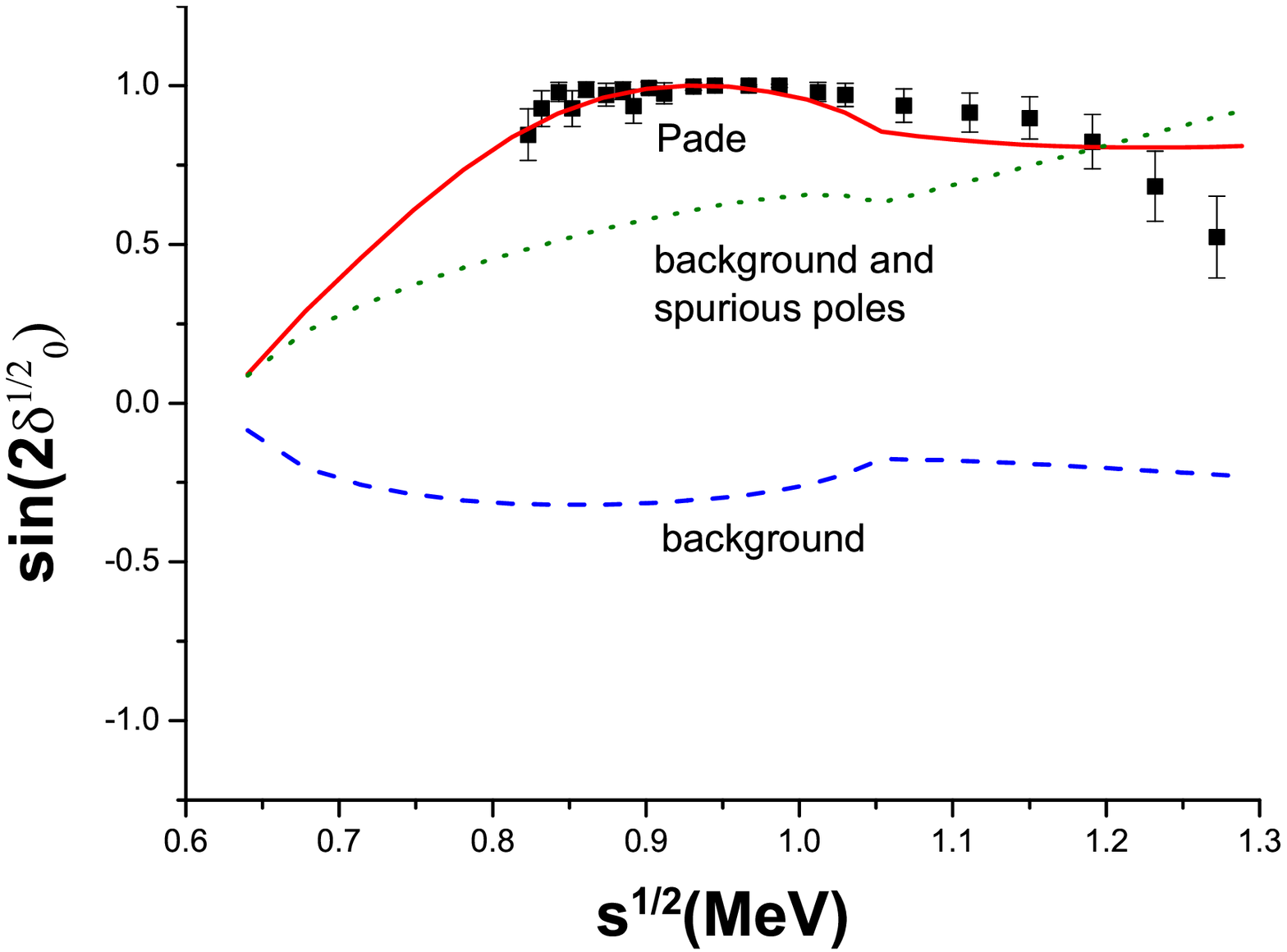}}%
\caption{\label{isospin-1} Various contributions to
$\cos2\delta^{1/2}_0$ and $\sin2\delta^{1/2}_0$ from the $\kappa$
pole, other spurious poles, left hand cut and right hand cut in
isospin $1\over 2$  S
wave $\pi K$ scattering. Here `spurious poles' means all poles other than the $\kappa$ pole.}%
\end{center}%
\end{figure}

Since the scattering amplitudes from chiral perturbation theory
only satisfy unitarity perturbatively and can not predict the
physical resonances by itself. We may restore unitarity by
constructing the  [1,1] Pad\'e approximant for the partial wave
amplitudes of $\pi K$ scattering,
\be\label{Pade amplitudes}%
T^{[1,1]}(s) =  {T_2(s)\over { 1 - {T_4(s)\over T_2(s)} }}\ .
\ee%
The [1,1] Pad\'e approximant given above satisfies elastic
unitarity if perturbative amplitudes satisfy the elastic unitarity
order by order,
\be
\mathrm{Im}{T^{[1,1]}(s)} = \rho(s)|T^{[1,1]}(s)|^2\ .%
\ee
 We take one set of values of the low-energy constants $L_r^i(\times 10^{3})$
from Ref.~\cite{Meissner}:
\bqa\label{low-energy constants}%
&&L_r^1 = 0.65\pm 0.28\ ,\ %
L_r^2 = 1.89\pm 0.26\ ,\ %
L_r^3 = -3.06\pm 0.92\ ,\nonumber \\ %
&&L_r^4 = 0.0\pm 0.5\ , \ %
L_r^5 = 2.2\pm 0.5\ , \nonumber \\ %
&&L_r^6 = 0.0\pm 0.3\ , \ %
L_r^8 = 1.1\pm 0.3\ .%
\eqa%
As an educative example, we use these values of low-energy
constants to study the singularity structure of the $S$ matrix of
Pad\'e approximation in the complex $s$ plane.
Poles of the $S$ matrix in the first and second sheet  in both
$I={3\over 2}$ and $I={1\over 2}$ channels are listed  in
table~\ref{table-1}. Since the $S$ matrix contains a circular cut
at $|s|=(m_K^2-m_{\pi}^2)$, in the table we indicate explicitly
whether the
pole locates inside or outside the  circular cut. %
\begin{table}[bt]%
\centering\vspace{-0.cm}%
\begin{tabular}{|c|c|c|c|c|c|}%
\hline%
I J            & Poles     & $\sqrt{s_p}$(GeV)         & Res[S($s_p$)]            \\%
\hline%
${3\over 2}$ 0 & Resonance & 0.23486 + 0.082077 i (inside)     &  0.0738734 - 0.00809522 i \\%
\cline{2-4}%
               & SPSR      & 0.131848 + 0.110795 i (inside)    &  -0.0115873 - 0.01906 i   \\%
\cline{2-4}%
               & SPSR      & 2.41263 + 1.477 i         &  -18.4928 + 5.16611 i     \\%
\hline%
${1\over 2}$ 0 & Resonance & 0.759984 + 0.297399 i     & -0.144373 - 0.613777 i    \\%
\cline{2-4}%
               & Resonance & 0.083424 + 0.449582 i (inside)     & 0.00286773 - 0.0699666 i  \\%
\cline{2-4}%
               & SPSR      & 0.0610924 + 0.0188401 i (inside)  & 0.0891571 + 0.673712 i    \\%
\cline{2-4}%
               & SPSR      &  0.299636 + 1.24856 i     & -2.18809 - 4.04587 i      \\%
\cline{2-4}%
               & SPSR      &  0.152309 + 0.343659 i (inside)   & 0.0634636 - 0.133458 i    \\%
\hline%
\end{tabular}%
\caption{\label{table-1}Resonances and spurious physical sheet
resonances (SPSR) predicted by the [1,1] Pad\'e approximant of
$\pi K$ scattering on the complex $s$ plane in both $I={3\over 2}$
and $I={1\over 2}$ channels  using the values from
Eq.~(\ref{low-energy constants}). The pole
position $\sqrt{s_p}\equiv M+i\Gamma/2$.}%
\end{table}%
From table~\ref{table-1}, we can find that the [1,1] Pad\'e
approximation predicts the existence of $\kappa$ resonance in
$I={1\over 2}$ channel, $M+i\Gamma/2 = 0.759984 + 0.297399 i$GeV,
but it also predicts many spurious poles. The effects of these
spurious poles are not always small. Especially in the $I={3\over
2}$ channel, the phase shift should have been given entirely by
the background (subtraction constant + cut integrals)
contributions, but in the Pad\'e approximant it is dominantly
contributed by the spurious pole contributions (see the figure for
$\cos 2\delta^{3/2}$ in fig.~\ref{isospin-3}). Various
contributions of physical poles, spurious poles, left hand cut
(including the circular cut) and right hand cut are clearly shown
in figs.~\ref{isospin-3} and \ref{isospin-1}. Beside the problems
mentioned above, Pad\'e approximation also fails to give the
correct $s$ dependence at $s=0$.~\cite{XZ03} Even though the
related studies have made remarkable successes in, for example,
correctly predicting various physical poles,\cite{Pade} we
conclude that such a unitarization approximation contains apparent
shortcomings, especially in the isospin 3/2 channel.

\section {A New Approach of Unitarization }\label{NAoU}

It is therefore necessary to find an alternative approach bridging
correctly the $S$ matrix theory and perturbation theory. It will
be the main purpose of this section. In here we generalize the
discussion made in Ref.~\cite{Ztalk03} to the general case of
unequal mass scattering and make a more complete analysis to the
new unitarization scheme.
\subsection{Simple $S$ Matrices}\label{SSM}

\begin{figure}%
\begin{center}%
\mbox{\epsfxsize=60mm\epsffile{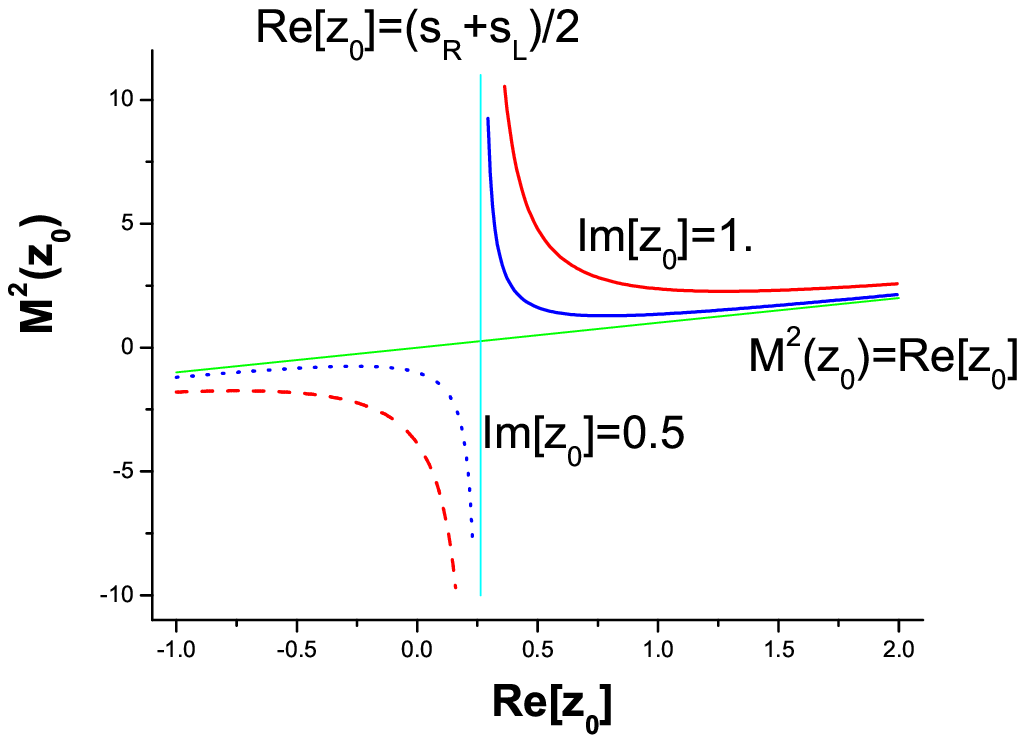}}%
\mbox{\epsfxsize=60mm\epsffile{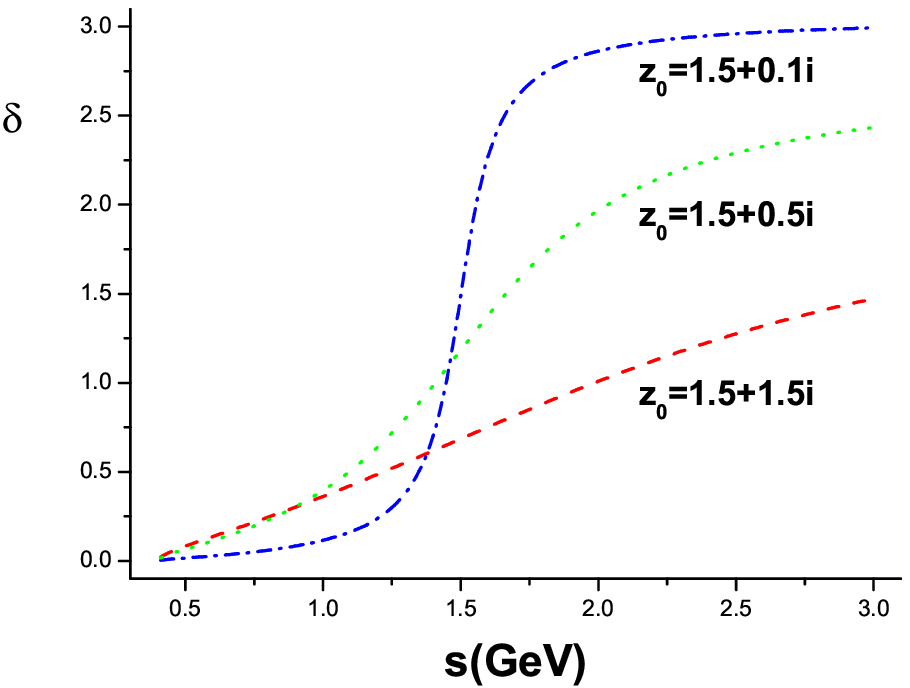}}%
\caption{\label{r0 function} The left figure shows $M^2(z_0)$ as a
function of $\mathrm{Re}[z_0]$, for two choices of
$\mathrm{Im}[z_0]$. The right figure gives some examples of
resonances and their contribution to phase shift.}
\end{center}%
\end{figure}

Single channel unitarity of the $S$ matrix tells us the following
identity:
 \bqa\label{GUE} {\cos^2 2\delta(s) } + {\sin
^22\delta(s) } \equiv 1\ , \eqa which is the generalized unitarity
relation and holds on the entire complex $s$ plane. A physical $S$
matrix is very complicated since it has various poles and cuts. We
overleap this complicated $S$ matrix and at this moment only
consider some simple circumstances. The word ``simple" means that
the $S$ matrix contains either only one pole or zero on the real
axis, or a pair of conjugated poles on the second sheet of the
complex $s$ plane. Meanwhile the simple $S$ matrices do not
contain cut contributions from those dispersion integrals in
Eq.~(\ref{dispersion-relations3}).
It is not difficult to find solutions of these simple $S$ matrices
by solving Eq.~(\ref{GUE}) and the solution for each kind of $S$
matrix is unique:
\begin{enumerate}
\item A virtual state pole at $s_0$ with $s_0$ real.
The solution to the scattering amplitude is,
 \bqa\label{ReTV}
\mathrm{Re_RT}(s) &=& \frac{s\,{\sqrt{\left( {s_R} - s_0 \right)
\,
        \left(s_0 -{s_L}  \right) }}}{
    \left( {s_R} - {s_L} \right) \,
    \left(s-s_0   \right) }\ ,\nonumber \\%
\mathrm{Im_RT}(s) &=&  \frac{\left( s -{s_R} \right) \,
      \left( {s_L} - s_0 \right) }{\rho (s)\left(
         {s_R} - {s_L} \right) \,
      \left( s-s_0 \right)}\ .
\eqa
 Consequently, the scattering length is
\bqa
 a(s_0)=\frac{2\,{\sqrt{{s_R}}}}{{s_R} -
    {s_L}}{\sqrt{\frac{s_0-{s_L}}
        {{s_R} - s_0}}}\ ,
\eqa
and the $S$ matrix can be expressed as, %
\bqa\label{a virtual state S matrix}%
S(s)=\frac{1+{i\rho }(s){s\over{s-{s_L}}}
{\sqrt{\frac{s_0-{{s}_L}}{{s_R}-s_0}
}}} {1-{i\rho }(s){s\over{s-{s_L}}} {\sqrt{\frac{s_0-{{s}_L}}{{s_R}-s_0} }}}\ .%
\eqa%
If there is a bound state at $s_0$, then ${\mathrm{Re_R}T}$ in
Eq.~(\ref{ReTV})  change sign. Besides, a bound or
virtual state exists only when $s_L<s_0<s_R$.%
\item A pair of resonances at $z_0$ (having positive imaginary part) and $z_0^*$:
The solution for scattering amplitude is,
\bqa%
\mathrm{Re_RT}(s) &=& \triangle(z_0)\mathrm{Re}[z_0 \rho (z_0)]
  \frac{s({M^2}(z_0)-s)}{(s-z_0)(s-{z_0^*})}\ ,\nonumber \\%
\mathrm{Im_RT}(s) &=& \triangle
(z_0)\mathrm{Im}[z_0]\frac{({s^2}+A(z_0)s+B(z_0))}
    {\rho (s)(s-z_0)(s-{z_0^*})}\ ,
\eqa%
and the scattering length is,
\bqa%
a(z_0)=\triangle (z_0)\mathrm{Re}[z_0\,\rho
(z_0)]\frac{2{\sqrt{{s_R}}}\,
    \left( {M^2}(z_0)-{s_R}\right)}{\left( {s_R} -z_0 \right) \,
    \left( {s_R} - {z_0^*} \right) }\ ,
\eqa%
where
\bqa%
\triangle(z_0) &=& \frac{\mathrm{Im}[z_0]}
  {{\mathrm{Im}[z_0]}^2 +
    {\mathrm{Re}[z_0\rho (z_0)]}^2}\ ,\nonumber \\
{M^2}(z_0) &=& \mathrm{Re}[z_0] + \frac{\mathrm{Im}[z_0]\,
     \mathrm{Im}[z_0\,\rho (z_0)]}{\mathrm{Re}[
     z_0\,\rho (z_0)]}\ ,\nonumber \\%
A(z_0) &=& - 2\mathrm{Re}[z_0]+\frac{\mathrm{Im}[z_0^2{\rho
(z_0)}^2]}
   {\mathrm{Im}[z_0]}\ ,\nonumber \\%
B(z_0) &=& {|z_0|}^2
  \left( 1 - \frac{\mathrm{Im}[z_0{\rho (z_0)}^2]}
     {\mathrm{Im}[z_0]} \right)\ .
\eqa%
The $S$ matrix can be expressed as:%
\bqa\label{a pair resonaces S matrix}%
S(s) = \frac{{M^2}(z_0)-s + i\rho(s)s G }
  {{M^2}(z_0)-s - i\rho(s)s G}\ ,
\eqa%
where
\bqa%
G = \frac{\mathrm{Im}[z_0]}{\mathrm{Re}[z_0\,\rho (z_0)]}\ .
\eqa%
\end{enumerate}
Analysis reveals interesting properties of  $M^2$ as a function of
$\mathrm{Re}[z_0]$ for fixed $\mathrm{Im}[z_0]$, as shown in
fig.~\ref{r0 function}. The inclined straight line corresponds to
$M^2=\mathrm{Re}[z_0]$ which is the asymptotic line of $M^2(z_0)$
when $\mathrm{Re}[z_0]\rightarrow \infty$. The vertical 
line corresponds to $\mathrm{Re}[z_0]=(s_R+s_L)/2$ and it is
another critical line on which the phase shift $\delta(s)$ from a
pair of resonances is $\pi/ 2$ when $s\rightarrow \infty$. On the
right hand side of the line, with the increase of $s$, the phase
shift of the resonances can get larger than $\pi/ 2$ whereas on
left hand side of the line, the phase shift of resonances can
never reach $\pi/ 2$. Fig.~\ref{r0 function} also gives some
examples of narrow and broad resonances and their contributions to
the phase shift. Each resonance gives phase shift a positive
contribution and the phase shift increases when $s$ increases. A
narrow resonance will give phase shift a sharp rise and behaves
much like an ordinary Breit--Wigner resonance, but a broad
resonance can only give phase shift a slow rise.

\subsection{A Pedagogical Analysis to a Toy Model Description of Resonances}
\begin{figure}%
\begin{center}%
\mbox{\epsfxsize=70mm\epsffile{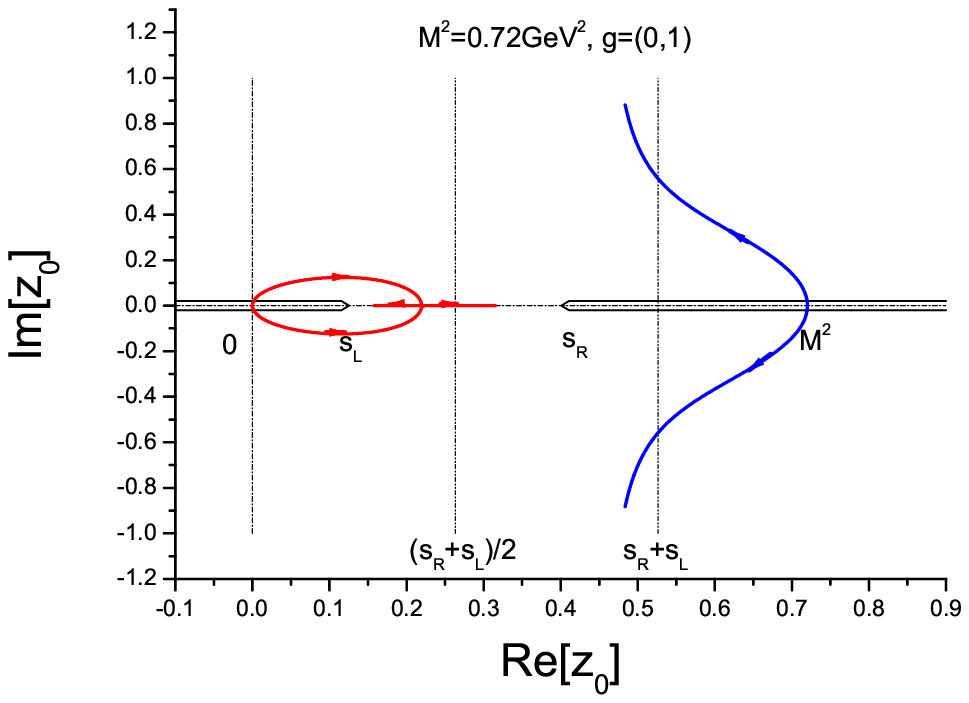}}%
\mbox{\epsfxsize=70mm\epsffile{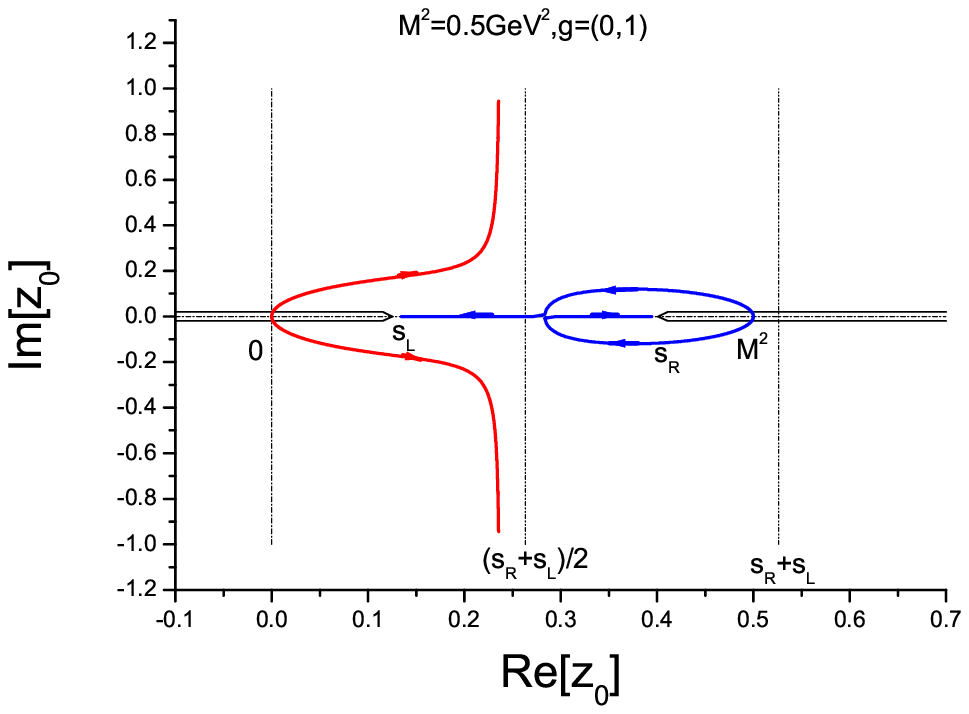}}%
\caption{\label{two pairs of resonaces} The traces of two pairs of
resonances from Eq.(\ref{two pairs of resonaces S matrix}) with
the increase of $g$ for different $M^2$. We give two typical
figures: the left figure for $M^2>(s_R+s_L)$, and the right figure
for $M^2<(s_R+s_L)$.}
\end{center}%
\end{figure}

A very simple but frequently used parametrization form of $S$
matrix to fit the position of a physical resonance is the
following,
\bqa\label{two pairs of resonaces S matrix}%
S(s) = \frac{M^2-s + i\rho(s)g }
  {M^2-s - i\rho(s)g}\ .
\eqa%
where $\rho(s)$ is the kinematic factor. For equal mass like the
$\pi\pi$ scatterings, $S$ matrix as described by Eq.~(\ref{two
pairs of resonaces S matrix}) usually contains a virtual state and
a pair of resonance poles on the second sheet. But Eq.~(\ref{two
pairs of resonaces S matrix}) usually contains two pairs of
resonance poles for $\pi K$ scattering due to the mass difference
between $\pi$ and $K$. Fig.~\ref{two pairs of resonaces} shows the
traces of two pairs of resonance poles with the increase of $g$
for different $M^2$. We find that $M^2=(s_R+s_L)$ and
$M^2=(s_R+s_L)/2$ are two critical points for the two pairs of
resonance poles. When $g$ increases from zero, two pairs of
resonances will appear, one from the origin (due to the $1/s$
singularity hidden in the kinematic factor) and another from
($M^2$, 0). When $g$ increases to a certain magnitude, one pair of
resonances will reach the real axis and change to two virtual
states, and one of them will run upwards  $s_R$ whereas another
will run downwards $s_L$. If $M^2>(s_R+s_L)$, the left pair of
resonance poles generated from the origin will change into two
virtual states and the right pair of resonance poles will become
wider and wider but never reach the critical line
$M^2=(s_R+s_L)/2$, and if $M^2<(s_R+s_L)$ the way of motion of the
two pairs of resonance poles changes, as shown in Fig.~\ref{two
pairs of resonaces}.

Eq.~(\ref{two pairs of resonaces S matrix}) is a commonly used
parametrization form of $S$ matrix to fit resonances. But from the
above discussion, we know that Eq.~(\ref{two pairs of resonaces S
matrix}) is not a parametrization of $S$ matrix for one pair of
resonance poles, but usually for two pairs. Since one pair of
poles is below threshold $s_R$ or change into two virtual states
on the real axis (also below the threshold), the existence of such
poles will violate the validity of chiral expansions at low
energies and is therefore spurious. Actually the low lying poles
in here has the same origin as the single virtual state pole in
the case of equal mass scattering.~\cite{Ztalk03} They are both
generated from the kinematical singularity of $\rho(s)$.  Whether
the existence of such spurious poles strongly affects the
determination of another pair of resonance poles depends on their
contribution to phase shift or scattering length. For example if
we set the physical resonance pole position at $\sqrt{s}=0.8\pm
0.3i$GeV
then another pair of poles will locate at
$\sqrt{s}=0.431\pm 0.125i$GeV. The scattering length contributed
by  the two are: $a=2.18$GeV$^{-1}$ for $\kappa$ and
$a=4.30$GeV$^{-1}$ for the spurious pole!

At this moment the major reason to exclude Eq.~(\ref{two pairs of
resonaces S matrix}) is rather academic. However, in
sec.~\ref{MethodI} we will see that the experimental data also
exclude very likely the use of Eq.~(\ref{two pairs of resonaces S
matrix}).
\subsection{The Factorized $S$ Matrix and the Separable
Singularities}\label{um}

 For a general $S$ matrix, there can be many poles on the
complex $s$ plane. However we can always express the complicated
physical $S$ matrix into a product of many simple $S$ matrices:
\bqa\label{factorized}
 S^{phy}=S^{cut}\cdot \prod_i S^{p}_i \ ,
\eqa%
suppose we can find all the poles of $S^{phy}$. The pole
contributions, $S_i^p$, to $S^{phy}$ are parameterized using the
forms given in Sec.~\ref{SSM}. Now the only uncertainty remains is
 the cut contribution $S^{cut}$, which, by
construction, contains no poles but inherits all the cut structure
of the original $S^{phy}$. Hence $S^{cut}$
can be parameterized  as the following:%
\bqa\label{FS}
 S^{cut}=\exp[2i\rho(s) f(s)]\ ,
\eqa
 where $f(s)$ satisfies the following  dispersion
relation: %
\bqa\label{Ff} f(s) &=& f(s_0)+{(s-s_0)\over 2\pi i}
{\int_L}{{\mathrm{disc_L}f(z)\over{(z-s)(z-s_0)}}dz}\nonumber \\%
&+& {(s-s_0)\over \pi}
{\int_R}{{\mathrm{Im_R}f(z)\over{(z-s)(z-s_0)}}dz}\ ,
 \eqa
where $L$ denotes the left hand cut on the real axis and also the
circular cut, and $R$ denotes the right hand cut starting from the
second physical threshold to $\infty$.
 Apparently the above parametrization automatically
guarantees
single channel unitarity. Furthermore we get,%
\bqa\label{IMLR}
\mathrm{disc}f=\mathrm{disc}\{\frac{1}{2i\rho(s)}\log
  \left[S^{phy}(s)\right]\}\
  \eqa%
  on both $L$ and $R$.
 The Eq.~(\ref{IMLR}) is derived, from an important property of
$S^p_i$, that is $the$ $latter$ $does$ $not$ $contribute$ $to$
$any$ $discontinuity$ $of$ $f$. The reason follows: firstly we
have
 \bqa f(s)&=&\frac{1}{2i\rho(s)}\log
 \left[S^{phy}(s)/\prod_iS^p_i\right]\nonumber\\
 &=&\frac{1}{2i\rho(s)}\log
 S^{phy}(s)-\frac{1}{2i\rho}\sum_i\log S^p_i(s)\ .
\eqa
 The first equality in the above equation ensures that $f$
 contains no more singularity than those cuts $S^{phy}$ contains,
 since poles and zeros of the two $S$ matrices on the $r.h.s.$ of the first
 equality exactly cancel, by definition. From the second
 equality, one easily understands, by comparing with the expressions of $S^p_i$ in
 sec.~\ref{SSM}, that the $S^p_i$ contribution to $\mathrm{disc}f$
 vanishes everywhere on all cuts. Therefore we have
 Eq.~(\ref{IMLR}). Especially, when $s$ is on the real axis, Eq.~(\ref{IMLR})
 can be rewritten as,
  \be\label{discf}
 \mathrm{Im}_{L,R}f(s)=-{1\over
{2\rho(s)}}\log|S^{phy}(s)|\ ,
  \ee
 which is a consequence of  real
 analyticity.
On the right hand cut $R=[(M_K+M_\eta)^2,\infty)$,
Eq.~(\ref{discf}) can be further rewritten as an analytic
expression,
 \be\label{IMR}
 \mathrm{Im}_{R}f(s)=-{1\over
{2\rho(s)}}\log\eta(s)=-{1\over {4\rho(s)}}
\log(\frac{S_{11}S_{22}}{\mathrm{det}S})\ ,
 \ee
 where the subscripts 1 and 2 mean channel $\pi K$ and channel
 $\eta K$, respectively (of course $S_{11}\equiv S^{phy}$, and in Eq.~(\ref{IMR}) $S$ means the $2\times 2$
 $S$ matrix of  couple channel scatterings). In principle, the Eq.~(\ref{factorized}) works not only in
 the single channel region but also works in the inelastic
 region. However, it should be
 emphasized that all poles in our formulae are on the second
 sheet. On the other side, the phase shift above the second physical threshold is mainly influenced
 by the third sheet pole rather than the second sheet pole (this is true at
 least for narrow resonances). Our present scheme suffers
 from the lacking of both theoretical\footnote{In this paper we do not attempt to study a true
  couple channel problem by giving an appropriate parametrization form of $\eta$. The problem was
  investigated in Ref.~\cite{XZ2} but was not very successful yet.}
 and experimental knowledge on the inelasticity parameter,
 $\eta$. It is not difficult to imagine the worst situation one
 may encounter when we have only insufficient information on
 inelasticity: suppose in the inelastic region under concern there
 is no second sheet pole but only a third sheet pole. The latter
 however only shows its effects through Eq.~(\ref{IMR}). If we neglect
  the inelasticity effects due to our ignorance
 on $\eta$, we would have to fit the phase shift data $\delta_2$ using a second sheet
 pole.  But this is certainly wrong by the assumption!
 Fortunately, the situation just described is not possible to
 happen in  $\pi K$ scatterings. Because the $K\eta$ cut is very
 weak up to the $K\eta'$ threshold~\cite{CP00}, the third sheet
 pole and the second sheet pole should co-exist and the mass and
 the width of the two poles should be similar either.\footnote{This may be
 best illustrated by a couple channel Flatt\'e model.} Therefore, for the
$K^*(1430)$ pole we are going to introduce in the later fit we
bear in mind that there are some ambiguities associate with it as
discussed above but the problem should not be serious. In the
later fit we will also neglect the $K\eta'$ threshold effects
which gives some further uncertainties to our final results. But
the uncertainties should not be large either  because we only fit
the data below the $K\eta'$ threshold where the influence from
high energy cuts should be small and smooth.
The uncertainties related to the right hand cut and the
$K^*(1430)$, furthermore, should not waver our main conclusions on
the $\kappa$ resonance, which is of our major concern in this
paper, since the energy region where all these problems occur, is
rather far from the low energy region where the $\kappa$ pole
locates.

 Going back to Eq.~(\ref{FS}) again, it factorizes different singularities of the
 scattering $S$ matrix. Also, different contributions to the phase
 shift are additive:
 \be\label{FPCd}
\delta=\sum_i\delta^p_i+\delta_{BG}\ ,
 \ee
where \be \delta^p=\mathrm{Atan}\left[\frac{\rho(s)s
G[z_0]}{M^2[z_0]-s}\right] \ee for a resonance located at $z_0$
(and $z_0^*$) and \be\label{dV}
\delta^p=\mathrm{Atan}\left[\frac{\rho(s)s}{s-s_L}\sqrt{\frac{s_0-s_L}{s_R-s_0}}\right]
\ee for a virtual state located at $s_0$ ($s_L<s_0<s_R$). The
bound state contribution can be obtained by simply change the sign
of $r.h.s.$ of Eq.~(\ref{dV}). For the background contribution we
have
 \be
 \delta_{BG}=\rho(s)f(s)\ .
\ee
 Notice that the separation  of pole contribution and background
contribution is only a matter of convention. However, our
definition of poles and background contribution has the  advantage
-- as manifested in Eq.~(\ref{discf}) --
that it greatly simplifies the calculations on various cuts. That
will be elaborated in the  next section.

The approximation scheme in evaluating the cut contributions is to
approximate $S^{phy}$ in Eq.~(\ref{discf}) by $S^{\chi
\mathrm{PT}}$ on $L$,
 \be
 \label{discf'}
  \mathrm{disc}f_{ L}=\mathrm{disc}\{\frac{1}{2i\rho(s)}\log
  \left[S^{\chi\mathrm{PT}}(s)\right]\}\ .
  \ee
Since the region where the above discontinuities are being
estimated are far from the resonance region, we expect the chiral
expansion (and hence our approximation scheme) works reasonably
well here, at moderately low energies. Also the logarithmic form
of the cut contribution automatically regulates, and hence reduces
the effect of, the bad high energy behavior of the chiral
amplitudes. Actually the cut integrals only need one subtraction
(Various cut integrals are all subtracted at the $\pi K$ threshold
throughout this paper). From these facts, we argue that our
approximation scheme for evaluating cuts are reasonable, at least
qualitatively. The quality of the approximation can be tested
experimentally, as is done in the next section.
\section {Fits to $\pi K$ Scattering Processes}\label{fit}

 Our goal is to search for resonances and to study their
properties, in $\pi K$ scatterings, by analyzing the phase shift
data. For this purpose, it is very important to first estimate
various background contributions, which are originated from
various cut integrals and the subtraction constant appeared in
Eq.~(\ref{Ff}) and (\ref{FPCd}). The correct understanding to the
background contributions is of course greatly helpful in the
attempt to a precise determination to the pole parameters. It is
even vital in answering the question whether there exists a broad
resonance, like the $\kappa$ resonance under debate. Since the
contribution from the background  and the contribution from  the
broad resonance can be rather similar, the lacking of the
knowledge on the former can even lead to completely misleading
prediction to the latter. It is therefore necessary to study
carefully the contribution of each term in  Eq.~(\ref{Ff}). One of
the main differences between the contribution from the background
and the contribution from a broad resonance is their contribution
to the scattering length. As we will show later that to evaluate
the scattering length parameter $a_0^{1/2}$ generated from fits is
very helpful in clarifying several issues, including the important
question whether $\kappa$ exists or not.
\subsection{The Estimation to the Background
Contributions}\label{BGc}

Our method of estimating the `left hand cut' contribution is to
substitute $S^{phy}$ in Eq.~(\ref{discf}) by $S^{\chi PT}$. The
background contribution from the left hand cut is then evaluated
by calculating the left hand cut integral in Eq.~(\ref{Ff}). The
once subtracted  integral is convergent and the integration is
formally performed from $-\infty$ to $s_L$ on the real $s$ axis.
However, since there is no reliable method to estimate correctly
the contribution from large negative $s$ region, it is more
appropriate to truncate the integration along the negative real
$s$ axis at $\Lambda^2_L$. This introduces an additional cutoff
parameter and we will test the dependence of the fit results on
this parameter.

The discussions made above clearly shows how approximations enter
into our scheme. It is necessary to first justify our
approximation scheme.
 Fortunately  it is possible to test whether it
is reasonable to use $\chi $PT to calculate the background
contributions. Because
 in the  I=3/2 channel there are only background contributions.
So we can compare directly the background contributions from our
approximation with experiments. We have already demonstrated in
sec.~\ref{pade} that the Pad\'e approximation gives poor
description in this channel to the background contribution. Here
we compare $\mathrm{Im}_{L}f(s)$ both using $S^{\chi \mathrm{PT}}$
and the [1,1] Pad\'e $S$ matrix, as shown in figs.~\ref{Imf
function3} and \ref{Imf
function3'}.\footnote{\label{fig7new}Various theoretical
estimations on $\delta^{3/2}_0$ in fig.~\ref{Imf function3'} are
calculated using the following formula:\begin{eqnarray*} f(s) &=&
{(s-s_A)\over
2\pi i} {\int_L}{{\mathrm{disc_L}f(z)\over{(z-s)(z-s_A)}}dz}\ ,%
 \end{eqnarray*}
where $\mathrm{disc_L}f(z)$ are estimated using the tree level, 1
loop $\chi$PT results and the Pad\'e result. The Adler zero $s_A$
for the Pad\'e result is taken as the tree level $\chi$PT result
on $s_A$. Notice that in this way the spurious poles'
contributions to the phase shift are absent. } We find once again
that the Pad\'e approximation result is poorer in reproducing the
phase shift data comparing with the $\chi$PT result after
adjusting the cutoff parameter within reasonable ranges.   Notice
that throughout this paper we do not attempt to make the global
fit by varying those $L_r^i$ parameters since the $L_r^i$
parameters are in principle threshold parameters. However, several
different results on the low energy constants found in the
literature are tested and it is found that the influences from the
different choices to the final fit results are very small. From
fig.~\ref{Imf function3'} disagreement between the theoretical
calculation  based on $\chi$PT  and the experimental data is
found. Better agreement can be achieved only by allowing
$a_0^{3/2}$ to be much larger than its $\chi$PT value.
Nevertheless it is verified that the problem in the I=3/2 channel
has rather small influence to the $\kappa$ pole problem. Further
discussions on this point will be given later.
\begin{figure}%
\begin{center}%
\mbox{\epsfxsize=70mm\epsffile{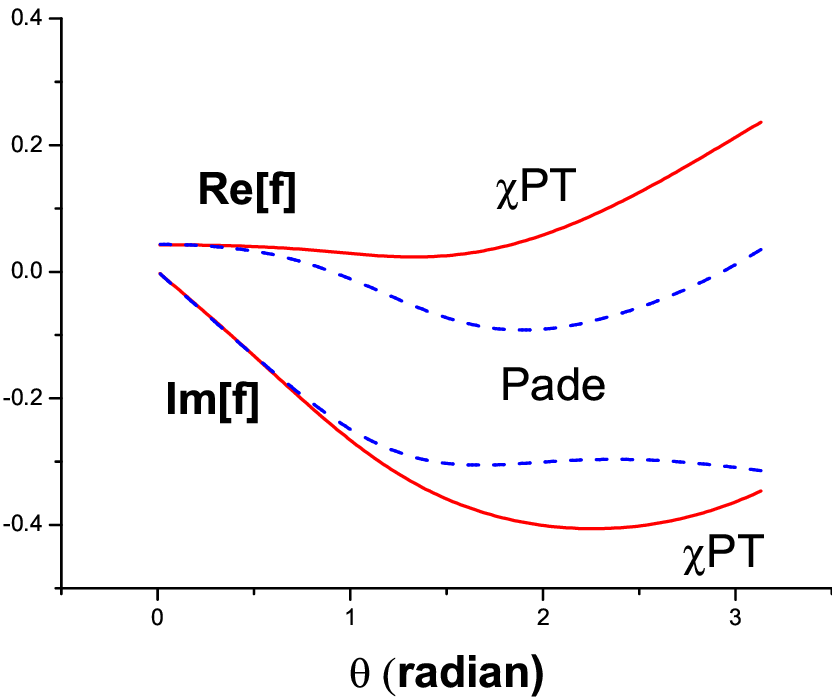}}%
\mbox{\epsfxsize=70mm\epsffile{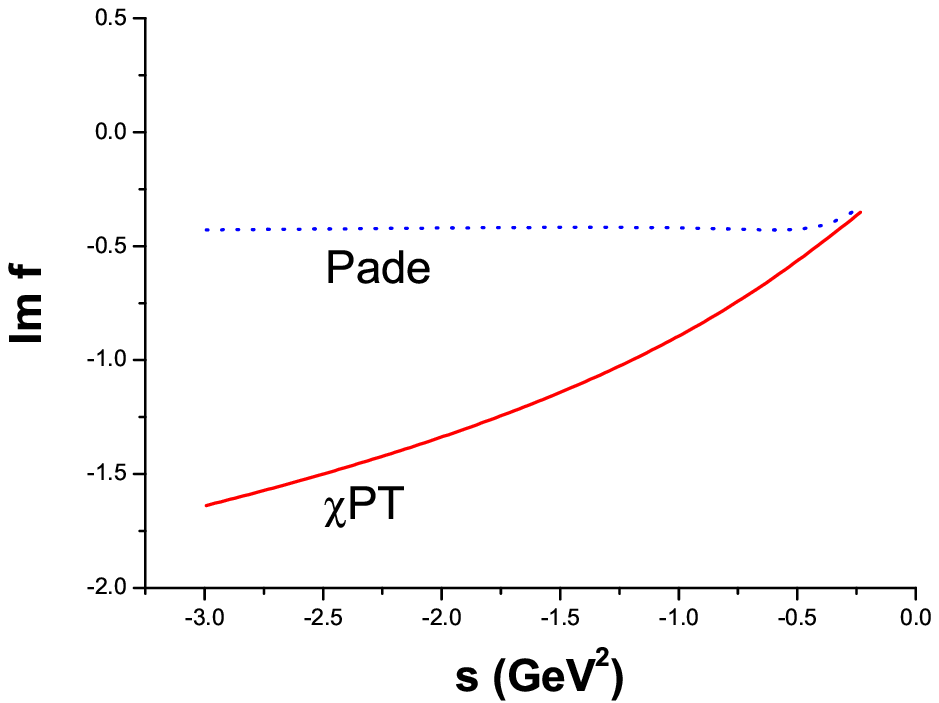}}%
\caption{\label{Imf function3}Various results on
$\mathrm{Im}_{L}f(s)$ or $f$ from $\chi$PT and Pad\'e
approximation in I=3/2 $s$ wave $\pi K$ scattering. Left: outside
the circular cut; right: on $L=(-\infty,-(M_K^2-M_\pi^2)]$. The
$L^i_r$ constants are taken from Eq.~(\ref{low-energy
constants}).}
\end{center}%
\end{figure}%
\begin{figure}%
\begin{center}%
\mbox{\epsfxsize=90mm\epsffile{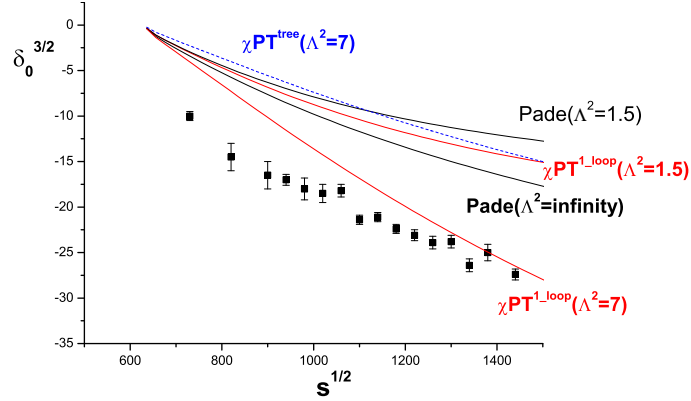}}%
\caption{\label{Imf function3'}Theoretical results on
$\delta_0^{3/2}$ versus the data~\cite{Estabrooks}. The dashed
line corresponds to the tree level $\chi$PT result. Other lines
are distinguished from the labels in the figure. See
footnote~\ref{fig7new} for more explanations.}
\end{center}%
\end{figure}%
\begin{figure}%
\begin{center}%
\mbox{\epsfxsize=70mm\epsffile{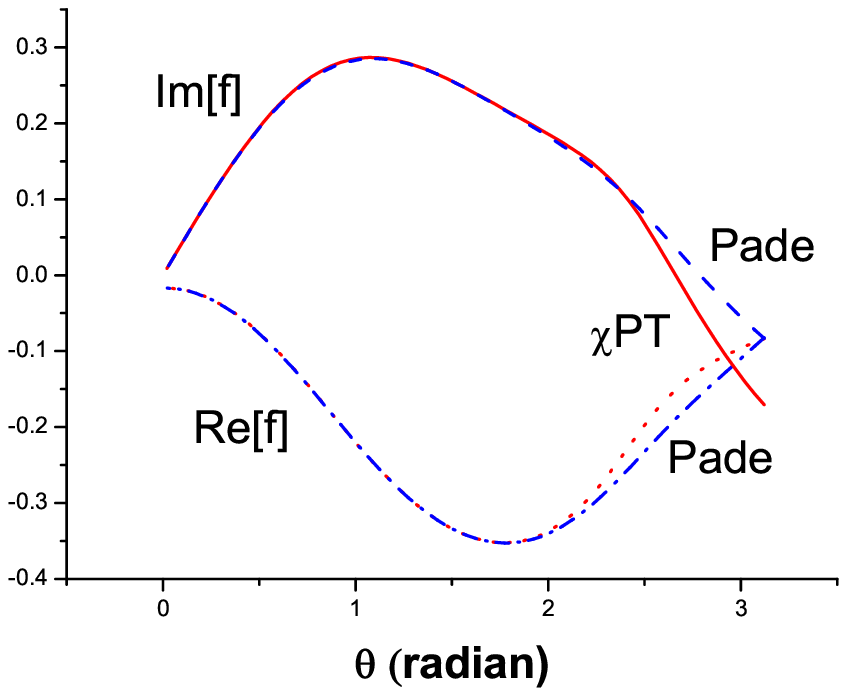}}%
\mbox{\epsfxsize=70mm\epsffile{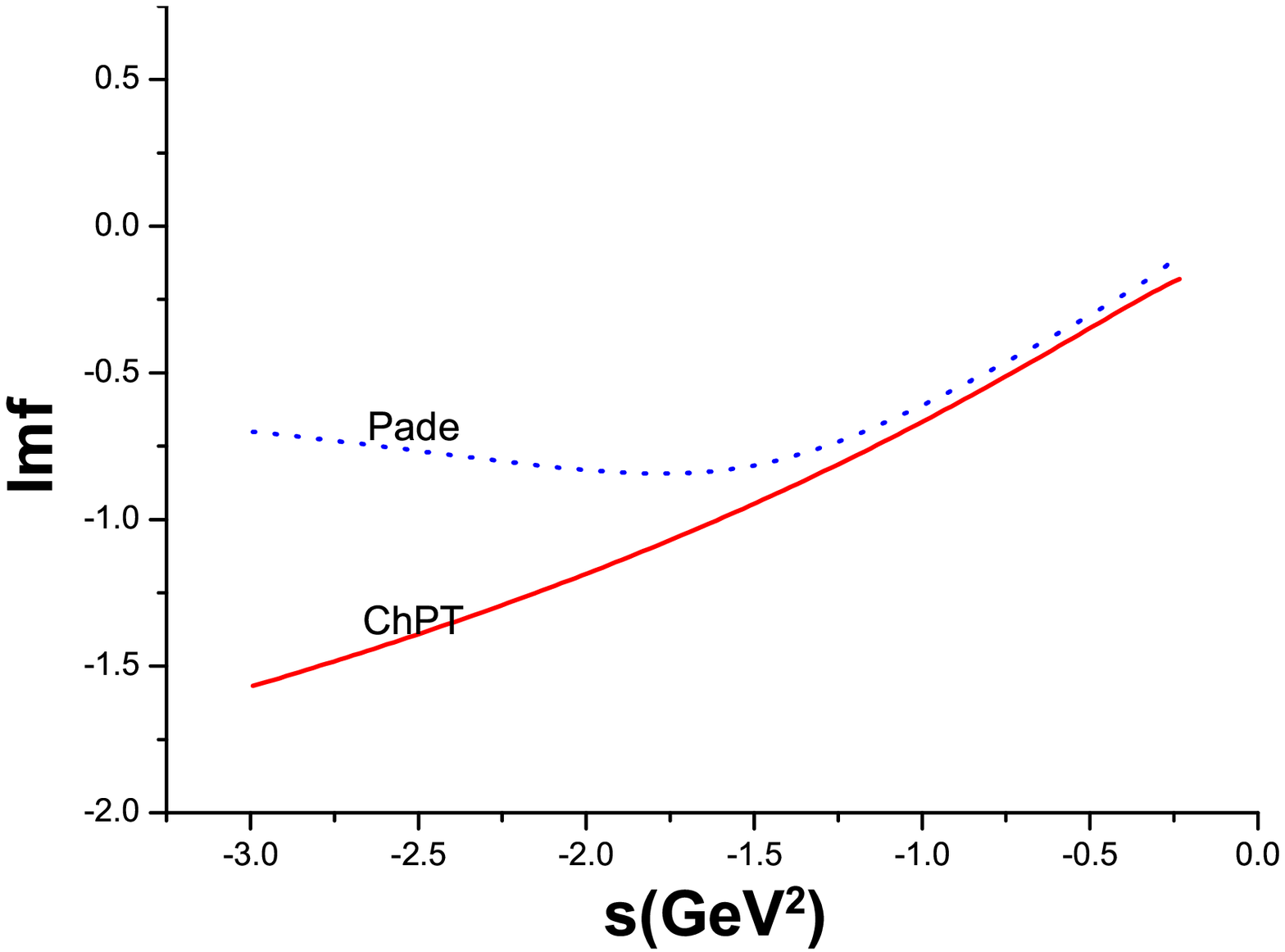}}%
\caption{\label{Imf function1} Various results of $f$ and
$\mathrm{Im}_{L}f(s)$ from $\chi$PT and Pad\'e approximation in
the I=1/2  channel. Left: outside the circular cut; right: on
$L=(-\infty,-(M_K^2-M_\pi^2)]$.}
\end{center}%
\end{figure}%

In the I=1/2 channel a direct check on the background contribution
is impossible since this is the channel where we will test the
existence of the $\kappa$ resonance. However from the experience
in the I=3/2 channel we expect that the $\chi$PT results on
$\mathrm{Im}_{L}f(s)$ works also well, at least qualitatively. It
is worth pointing out that the results from the [1,1] Pad\'e
approximant are rather similar to the $\chi$PT results in the
I=1/2 channel, except at large negative s region. Therefore we
will use the $\chi$PT prediction on $\mathrm{Im}_{L}f(s)$.
  In the  I=1/ 2 channel, there exists also the right hand cut from
$s=(M_K+M_{\eta})^2$ but it is very weak till another threshold
 $s=(M_K+M_{\eta'})^2$ is reached. For the right hand
cut, $\chi $PT result is totally misleading because it violates
single channel unitarity in the physical region. On the other
side, the Pad\'e result maintains single channel unitarity but may
give a too large contribution to the inelasticity parameter,
starting from $(M_K+M_\eta)^2$. Since experiments suggest that the
inelasticity is very small till $(M_K+M_{\eta'})^2$ and the latter
is already very far from the low energy region we are interested
in, we during the fit simply set the right hand cut contribution
being vanish. As is already discussed in sec.~\ref{um}, this
approximation may have some  effects to the determination of the
$K(1430)$ resonance but should have negligible influence to the
$\kappa$ resonance.
\subsection{The Numerical Analysis and Data Handling}
The two experiments on $K\pi$ scatterings we will make use of are
from the LASS Collaboration~\cite{ASton} and Estabrooks et
al.~\cite{Estabrooks}. The latter gives the phase shift data   in
the I=1/2  channel for $\sqrt{s}=0.73 - 1.3$GeV and in the I=3/2
channel for $\sqrt{s}=0.73 - 1.72$GeV. It is found that the
scattering is purely  elastic up to 1.3GeV, i.e.,
$\eta_0^{1/2}=\eta_0^{3/2}=1$ in  this region. In the I=3/2
channel the inelasticity can be neglected in the whole energy
range.
 The experiment by the LASS Collaboration only
measures the $K^-\pi^+$ channel and hence only affords the
following combination of data:
 \be\label{a0phi0}
  A_0=a_0e^{i\phi_0}=T_0^{1/2}+\frac{1}{2}T_0^{3/2}=\frac{1}{2i}(\eta_0^{1/2}e^{2i\delta^{1/2}_0}-1)
  +\frac{1}{4i}(\eta_0^{3/2}e^{2i\delta^{3/2}_0}-1)\ ,
\ee
 for $\sqrt{s}=0.825 - 2.52$GeV. Notice that in the above equation the
 definition of $T$ matrix is different from previously used in
 this  paper.

 In this work our discussion will be  confined to the single channel
 approximation,
  though it is
 understood that our formalism in principle works also in the inelastic region.
The validity domain of single channel approximation is largely
expanded  in $\pi K$ scatterings since the second threshold, the
$K\eta$ channel opens very weakly in agreement with $SU(3)_f$
expectations, and so any inelasticity can be  neglected until one
reaches the $K\eta'$ threshold~\cite{CP00}. Therefore we in the
fit assume elasticity up to the $K\eta'$ threshold both in I=1/2
and 3/2 channels. As illustrated in the Introduction the main
purpose of this paper is to study whether there exists the
$\kappa$ resonance and, if exists, its properties. The
approximation of neglecting the inelasticity effects  will mainly
affect the well established $K^*(1430)$ resonance which is not our
main interest here.

Our strategy of making the fit follows: assuming two resonances,
one for $K^*(1430)$, another one for the $\kappa$ resonance under
investigation and the two complex poles  contribute four
parameters. There are another two parameters coming from the two
scattering lengths, i.e., $a_0^{1/2}$ and $a_0^{3/2}$ (or
equivalently, two subtraction constants). Therefore there are
totally 6 parameters in the fit. The first fit only make use of
the LASS data  up to $\sqrt{s}=1.43$GeV, which is about 20MeV
below the $K\eta'$ threshold and consist of 60 data points. We
call it {\bf Method I} hereafter. In the second fit we also
include data from Estabrooks et al.,  up to $\sqrt{s}=1.3$GeV in
the I=1/2 channel and up to $\sqrt{s}=1.43$ in the I=3/2 channel.
This will add another 41 data points, among them 24 come from
 the $\delta_0^{1/2}$ phase shift data and 17 from the
 $\delta_0^{3/2}$ phase shift data. We call this fit scheme {\bf Method II}
 hereafter. Treatment to various cuts are already illustrated in
 sec.~\ref{BGc}.

\subsection{The fit to the LASS data (Method I)}\label{MethodI}
This part of discussion is divided into two subsections: one is
the fit without considering further constraints from chiral
perturbation theory except when estimating the left hand cut
contributions. The second subsection respects the $\chi$PT
results, especially its predictions on the scattering length
parameters. Since what is involved here is the SU(3) version of
$\chi$PT and since there exists possible conflict between theory
and experiments, we think it is worthwhile to be cautious to make
the separate discussions.
\subsubsection{The fit without constraints from $\chi$PT}
 Using MINUIT, we perform the six parameters fit
($a_0^{1/2}$, $a_0^{3/2}$ and four pole parameters for  $\kappa$
and $K^*(1430)$) to the LASS data up to 1430MeV. Except for these
6 fit  parameters we have, as already stressed, one additional but
somewhat unpleasant parameter: the cutoff parameters responsible
for the truncation of left hand integrals in both  I=1/2 and 3/2
channels, denoted as $\Lambda_{L}^2$.\footnote{\label{5}We have
also carefully checked the situation when $\Lambda^2_L$ in the
I=1/2 channel and in the I=3/2 channel are different. The final
results are very similar to those presented in this paper.}
The fit results will of course depend on the cutoff parameter,
nevertheless the  dependence  is much weaker than that of
Ref.~\cite{XZ00} which is of course progressive. In
table~\ref{tab2}  we list various results obtained by
 varying $\Lambda^2_{L}$ 
 from which
 we draw the following conclusions:
 \begin{enumerate}
 \item The overall $\chi^2$ is rather stable against most changes
 of the cutoff parameters, and the fit prefers $\Lambda^2_{L}\simeq 1.5$GeV$^2$. This
is supported by an independent analysis to the  phase shift data
provided by Ref.~\cite{Estabrooks}. It should be emphasized that
these conclusions are obtained only when $a_0^{1/2}$ and
especially $a_0^{3/2}$ are treated as completely free in the fit.
If $a_0^{3/2}$
is confined to its $\chi$PT value, the best $\Lambda^2_{L}$ value
will be enhanced.
 \item The fit result on $a_0^{3/2}$ is rather sensitive to the cutoff parameter.
 Actually these sensitivity are only
 related to the cutoff parameter in the I=3/2 channel rather than  that in the I=1/2
 channel (see also footnote~\ref{5} and fig.~\ref{Imf function3'}).
 Remember that we are fitting the combined data now,
 therefore our method can somehow rather clearly distinguish
 different channel contributions to the LASS data.
 \item Except for the problem mentioned above, most other results are rather stable
 against the variation of $\Lambda_L^2$ in a reasonable
 range. This is even true when we confine $a_0^{1/2}$ and
 $a_0^{3/2}$ in the fit. See later text for more discussions.
\end{enumerate}
To save the already lengthy discussions we in the following will
only exhibit  results  for the fixed value of the cutoff
parameters: $\Lambda^2_{L}=1.5$GeV$^2$, unless otherwise stated.
 However we also carefully analyzed the uncertainties of all our major outputs
 induced by the uncertainties of the cutoff parameters
  and found out that the uncertainties are not magnificent.
\begin{table}[bt]
\centering\vspace{0.1cm}
\begin{tabular}{|c|c|c|c|c|c|}
\hline
$\Lambda^2_{L}$&$\chi^2_{tot}$&$a_0^{1/2}$&$a_0^{3/2}$&$M_\kappa$
&$\Gamma_\kappa$
\\ \hline
1&38.64&$0.292\pm 0.100$  &$-0.143\pm 0.006$&$573\pm 108$&$747\pm
438$
\\ \hline
1.5&38.35 &$0.284 \pm 0.089$ &$-0.129\pm 0.006$&$594\pm
79$&$724\pm 332 $
\\ \hline
2&39.44 &$0.278 \pm 0.078$ &$-0.118\pm 0.006$&$609\pm
58$&$711\pm 272 $
\\ \hline
2.5&41.30 &$0.274 \pm 0.087$ &$-0.109\pm 0.006$&$620\pm
61$&$703\pm 258 $
\\ \hline
3&43.55 &$0.271 \pm 0.129$ &$-0.102\pm 0.006$&$629\pm
81$&$697\pm 354 $
\\ \hline
3.5&45.96 &$0.268 \pm 0.077$ &$-0.096\pm 0.006$&$636\pm
40$&$693\pm 236 $
\\ \hline
5&53.18 &$0.262 \pm 0.068$ &$-0.082\pm 0.006$&$652\pm
30$&$685\pm 191 $
\\ \hline
10&71.40 &$0.254 \pm 0.062$ &$-0.060\pm 0.006$&$676\pm
24$&$673\pm 148 $
\\ \hline
$\infty$&120 &$0.242\pm 0.067$ &$-0.026\pm 0.007$&$716\pm
16$&$653\pm 136 $
\\ \hline
\end{tabular}
\caption{\label{tab2}Various fit results to the LASS data up to
1.43GeV obtained by varying the cutoff parameters of the left hand
integrals. There are totally 60 data points and 6 parameters.
Various results on $K^*(1430)$ are  similar and are not listed
here. All values of  mass parameters in the table are in units of
MeV.}
\end{table}

\begin{figure}%
\begin{center}%
\mbox{\epsfxsize=80mm\epsffile{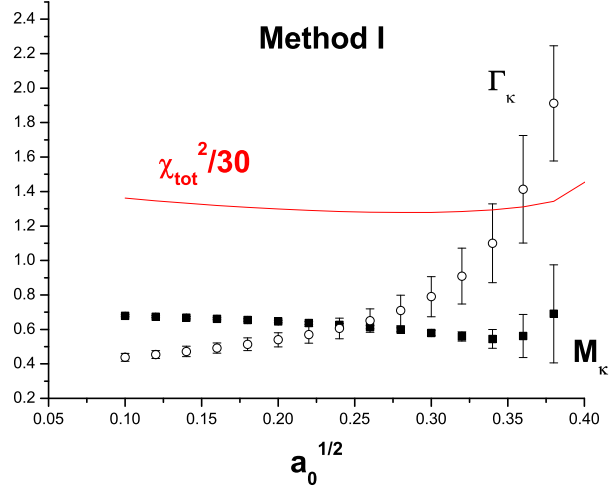}}%
\caption{\label{fig3a1}The dependence of the mass, width and their
error bars of the $\kappa$ resonance on $a_0^{1/2}$ (Method I).
The fits contain only 5 parameters. Taking
$\Lambda_L^2=1.5$GeV$^2$. }
\end{center}%
\end{figure}

 As is shown in table~\ref{tab2}, the global fit
prefers larger magnitudes of the scattering lengths than those
predicted by $O(p^4)$ $\chi$PT~\cite{Meissner}, which are:
$a_0^{1/2}=0.18\pm 0.02$, $a_0^{3/2}=-0.05\pm 0.02$. The present
fit gives however rather large error bars on $a_0^{1/2}$.
To further test the dependence of the $\kappa$ pole on $a_0^{1/2}$
we also perform the fits by varying $a_0^{1/2}$ while keeping it
fixed during each fit. The results are shown in fig.~\ref{fig3a1}.
In fig.~\ref{fig3a1} we have not shown the situation when
$a_0^{1/2}>0.4$ where the results become more and more unstable.
From fig.~\ref{fig3a1} we may have the impression that when
$a_0^{1/2}$ is roughly less than 0.35 the $\kappa$ resonance seems
to exist. If $a_0^{1/2}$ is greater than the value the existence
of the resonance becomes doubtful. This observation will be
further examined in the following discussions.

\begin{figure}%
\begin{center}%
\mbox{\epsfxsize=65mm\epsffile{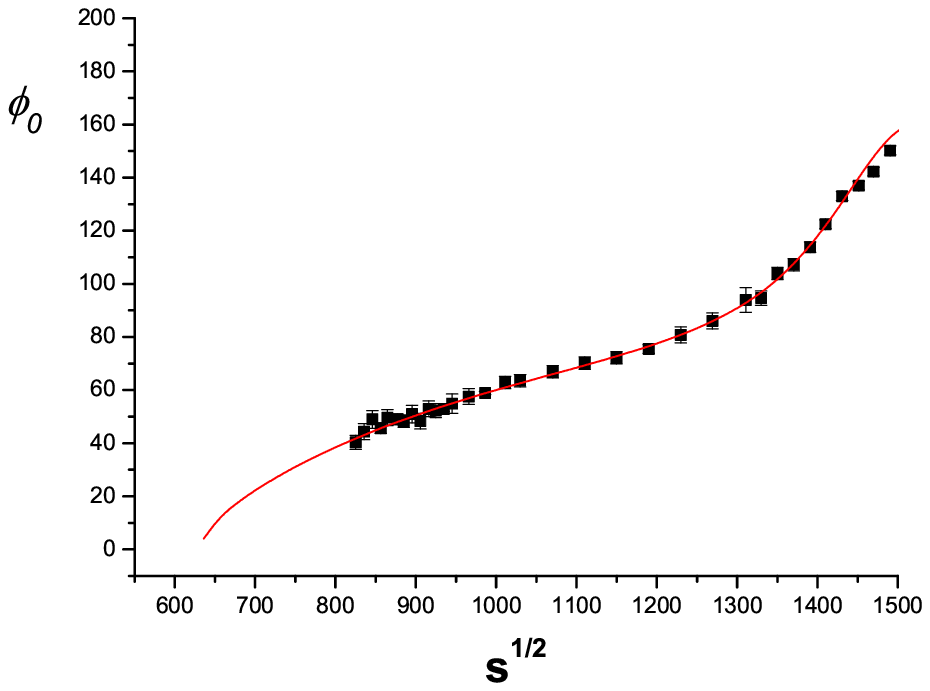}}%
\hspace{.15cm}
\mbox{\epsfxsize=65mm\epsffile{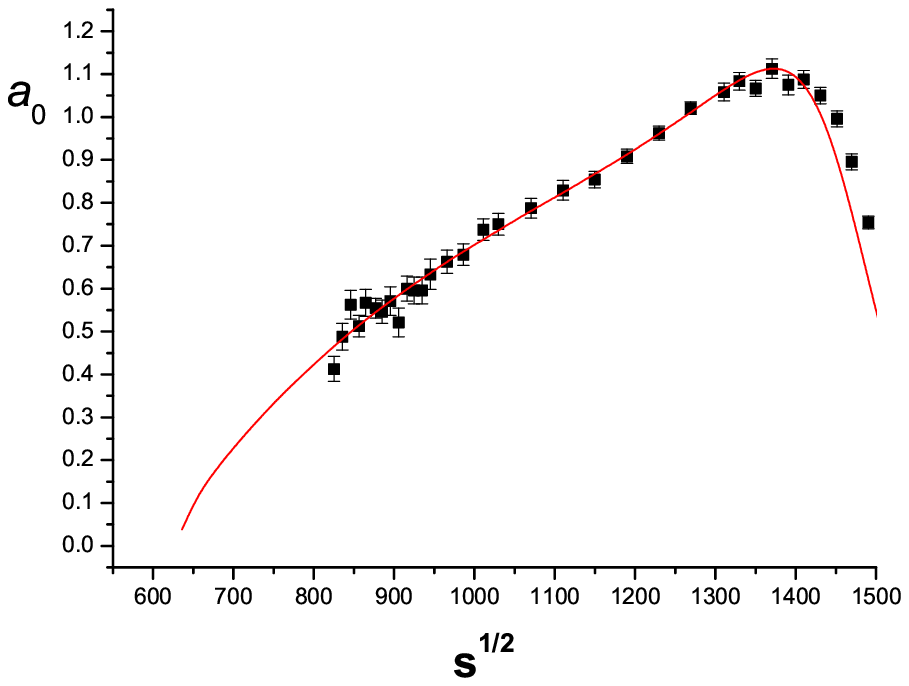}}%
\caption{\label{figphi0} Fit results (Eq.~(\ref{resI})) to
$\phi_0$ (left) and $a_0$ (right) using Method I. The data are
fitted only up to 1430MeV but are shown up to 1500MeV.
 }
\end{center}%
\end{figure}

\begin{figure}%
\begin{center}%
\mbox{\epsfxsize=70mm\epsffile{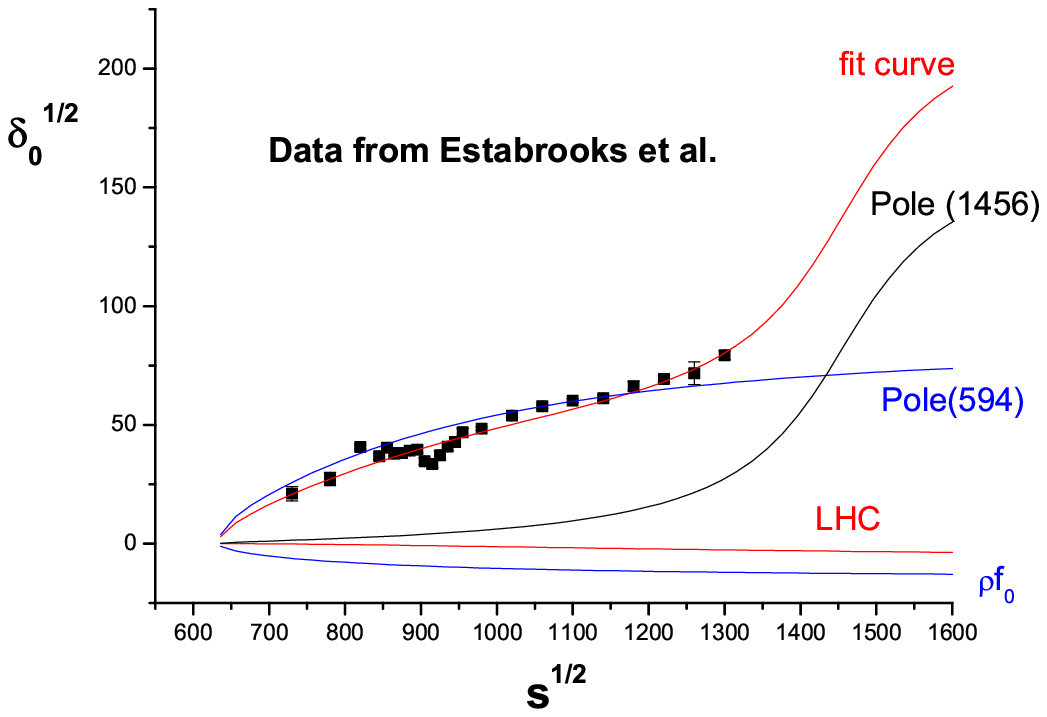}}%
\mbox{\epsfxsize=70mm\epsffile{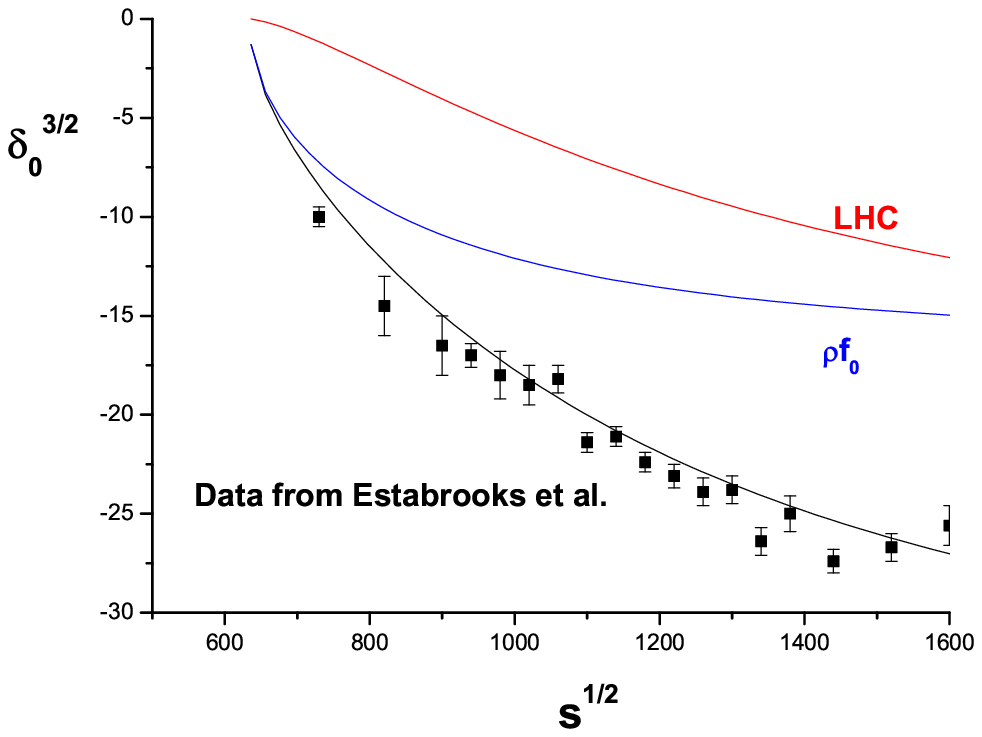}}%
\caption{\label{figdelta1} The comparison between the fit results
on the LASS data (Eq.~(\ref{resI})) and the data from Estabrooks
et al.~\cite{Estabrooks}.}
\end{center}%
\end{figure}

The full results of the six parameters fit to the LASS data are
the following:
 \bqa\label{resI}
&&\chi^2_{d.o.f.}=38.35/(60-6)\ ;\nonumber\\
&&M_\kappa=594\pm 79MeV\ ,\,\,\, \Gamma_\kappa=724\pm 332MeV\ ;\nonumber\\
&&a_0^{1/2}=0.284\pm 0.089\ ,\,\,\,a_0^{3/2}=-0.129\pm 0.006\ ;\nonumber\\
&&M_{K^*}=1456\pm 8MeV\ ,\,\,\, \Gamma_{K^*}=217\pm 31MeV\ .
 \eqa
 The corresponding fit results are plotted in fig.~\ref{figphi0}.
Since their is no separate data of I=1/2 and I=3/2 channels
provided by the LASS Collaboration, we also plot curves of
$\delta_0^{1/2}$ and $\delta_0^{3/2}$ from our fit results versus
the phase shift data from Estabrooks et al. in
fig.~\ref{figdelta1}. We stress again that in the present fit the
LASS data are combined data from both the I=1/2 and I=3/2
channels. From fig.~\ref{figdelta1} we find that the present
method can  rather clearly distinguish different contributions
from different channels.

 In order to answer the question whether $\kappa$ exists or not,
  we further freeze the $\kappa$ degrees of freedom  in the fit and  we get
  \bqa\label{resI'}
&&\chi^2_{d.o.f.}=63.67/(60-4)\ ;\nonumber\\
&&M_\kappa=--\ ,\,\,\, \Gamma_\kappa=--\ ;\nonumber\\
&&a_0^{1/2}=0.446\pm 0.006\ ,\,\,\,a_0^{3/2}=-0.130\pm 0.006\ ;\nonumber\\
&&M_{K^*}=1432\pm 3MeV\ ,\,\,\, \Gamma_{K^*}=314\pm 20MeV\ .
 \eqa
Comparing with the results in Eq.~(\ref{resI}) the
$\chi^2_{d.o.f.}$ given by Eq.~(\ref{resI'}) is increased by a
factor of 1.7. If this is  not enough to support the existence of
$\kappa$, the value of $a_0^{1/2}$ given by Eq.~(\ref{resI'}) is
too large comparing with the $\chi$PT value. While the fit value
of $a_0^{3/2}$ is about 4 $\sigma$ away from $\chi$PT result,
 $a_0^{1/2}$ as predicted by Eq.~(\ref{resI'}) is
14$\sigma$ away! It is actually easy to understand why the fit
program is forced to chose such a large $a_0^{1/2}$. Looking at
fig.~\ref{figdelta1}, when the contribution of the $\kappa$
resonance is withdrawn the subtraction constant has to be much
increased to fit data. Since the $\rho f_0$ contribution increase
faster at threshold than the $\kappa$ contribution and slower at
higher energies than the $\kappa$ contribution, this leads to a
larger $a_0^{1/2}$.

It is also worthwhile to  investigate the fit using Eq.~(\ref{two
pairs of resonaces S matrix}) to parameterize $\kappa$. The result
follows:
 \bqa\label{resIb}
&&\chi^2_{d.o.f.}=38.22/(60-6)\ ;\nonumber\\
&&M_\kappa=726\pm 8MeV\ ,\,\,\, \Gamma_\kappa=698\pm 19MeV\ ;\nonumber\\
&&a_0^{1/2}=0.403\pm 0.025\ ,\,\,\,a_0^{3/2}=-0.129\pm 0.006\ ;\nonumber\\
&&M_{K^*}=1453\pm 3MeV\ ,\,\,\, \Gamma_{K^*}=220\pm 18MeV\ ,
 \eqa
and the spurious resonance pole locates at (the central value):
$M=477$MeV, $\Gamma=94$MeV. Comparing with Eq.~(\ref{resI}), the
total $\chi^2$ given by Eq.~(\ref{resIb}) is very similar, but it
gives a too large $a_0^{1/2}$ comparing with the $\chi$PT
prediction. Therefore Eq.~(\ref{two pairs of resonaces S matrix})
is very likely to be excluded by the LASS data.
\subsubsection{The fit with constraints from
$\chi$PT}\label{adler}
 It is
sometimes found in the literature the discussions on the
constraint of the Adler zero on the scattering
amplitude.~\cite{Bugg03} In our scheme, it is possible to embed
this constraint into our parametrization form, similar to what is
done in Ref.~\cite{HXZ02}. However, the Adler zero automatically
emerges in our approach if the subtraction constant $f_0$ is
limited within certain range. This is because all the
resonance(and virtual state) $S$ matrices are real and less than 1
when $s_L<s<s_R$. On the contrary the cut integrals contribute a
factor larger than 1, therefore a $T$ matrix zero emerges  in the
right place when $f_0$ is confined to a certain range, which in
turn put some constraints on the magnitude of the scattering
length parameter itself. For example, within the range
$0<a_0^{1/2}<0.26$ (see fig.~\ref{fig3a1}) there exists a $T$
matrix zero in the region $s_L<s<s_R$, and when $a_0^{1/2}\simeq
0.20$ the zero locates in the place close to the one loop $\chi$PT
prediction. We make further fit by confining $a_0^{1/2}$ in the
region $0.18\pm 0.02$  and the results follow: \bqa\label{resIa}
&&\chi^2_{d.o.f.}=38.96/(60-6)\ ;\nonumber\\
&&M_\kappa=646\pm 7MeV\ ,\,\,\, \Gamma_\kappa=540\pm 42MeV\ ;\nonumber\\
&&a_0^{1/2}=0.2\ ,\,\,\,a_0^{3/2}=-0.128\pm 0.006\ ;\nonumber\\
&&M_{K^*}=1450\pm 5MeV\ ,\,\,\, \Gamma_{K^*}=232\pm 25MeV\ .
 \eqa
The Adler zero position is now at $s_A\simeq 0.245$GeV to be
compared with the 1--loop $\chi$PT value $s_A\simeq 0.233$GeV. The
necessity for the existence of the $\kappa$ resonance  may be best
illustrated by the fit when constraining $a_0^{1/2}$ to lie within
the range as predicted by $\chi$PT value, $a_0^{1/2}=0.18\pm 0.02$
and meanwhile freeze out $\kappa$. Under this situation $a_0^0$
will reach its upper value at 0.2 and the $\chi^2_{tot}\sim750$!
Comparing with the value of $\chi^2_{tot}$ in Eqs.~(\ref{resIa})
and (\ref{resI}), it clear demonstrates the necessity to include
the $\kappa$ resonance if $a_0^{1/2}$ is close to (or not much
larger than) its $\chi$PT value.

The next question  is to ask what would happen if we further
confine both $a_0^{1/2}$ and $a_0^{3/2}$ to their $\chi$PT value?
The answer is,
 \bqa\label{resIb'}
&&\chi^2_{d.o.f.}=127.6/(60-6)\ ;\nonumber\\
&&M_\kappa=655\pm 9MeV\ ,\,\,\, \Gamma_\kappa=549\pm 42MeV\ ;\nonumber\\
&&a_0^{1/2}=0.2\ ,\,\,\,a_0^{3/2}=-0.07\ ;\nonumber\\
&&M_{K^*}=1465\pm 5MeV\ ,\,\,\, \Gamma_{K^*}=258\pm 33MeV\ .
 \eqa
 Though the $\chi^2_{tot}$ is now much enhanced as comparing with
 the result given by Eq.~(\ref{resIa}), we are very much consoled
 by the nice agreement on the $\kappa$ pole between the two
 methods. This again suggests that though we fit the LASS data which is
 a  combined effect of both the I=1/2 channel and the I=3/2
 channel, the present approach can somehow
 clearly distinguish different contributions from different
 channels. Actually if we make a plot like fig.~\ref{figdelta1}
 we find that the fit curve in the I=1/2 channel still agrees  well with the data  whereas
 in the I=3/2 channel there exists rather large deviations. Technically,
 the large $\chi^2_{tot}$ appeared in Eq.~(\ref{resIb'}) can be
 reduced roughly  by half when increasing $\Lambda_{L}^2$ (or more precisely, increasing
 the cutoff parameter in the I=3/2 channel).
 This ambiguity with respect to the cutoff parameter will contribute, though not large,
  some uncertainties
 to the pole positions, which can be considered as the `systematic' error in our approach.

\subsection{The combined data fit (Method II)}

\begin{figure}%
\begin{center}%
\mbox{\epsfxsize=80mm\epsffile{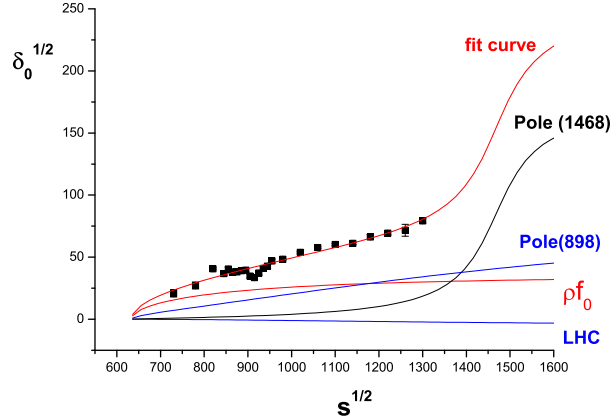}}%
\caption{\label{figd1C} The combined fit results
(Eq.~(\ref{resIII})) on $\delta_0^{1/2}$.}
\end{center}%
\end{figure}

\begin{figure}%
\begin{center}%
\mbox{\epsfxsize=80mm\epsffile{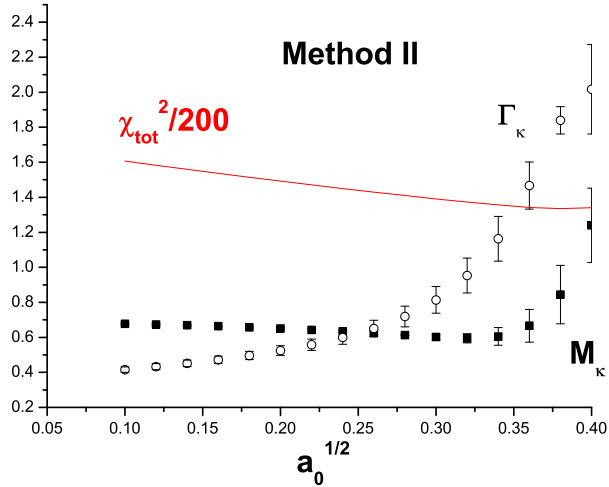}}%
\caption{\label{fig3a1AE}The dependence of the mass, width and
their error bars of the $\kappa$ resonance on $a_0^{1/2}$ (Method
II). Taking $\Lambda_{L}^2=1.5$GeV$^2$.}
\end{center}%
\end{figure}

 The data from Estabrooks et al. contain some problems. This
 can be clearly seen from fig.~\ref{figdelta1}. The  dip structure of
 the phase shift data around 920MeV apparently leads to $\frac{d\delta}{ds}<0$.
 According to our analysis, it cannot be explained by
 the smooth background contributions. If the dip structure were true
 it unavoidably leads to very complicated pole structures and causality violation conclusion.
 The dip
 structure actually gives a large contribution to the total
 $\chi^2$ in the combined fit to both the LASS data and the data
 from Estabrooks et al.. Nevertheless we still make the combined
 fit using the method as described previously. The result for six
 free parameters fit
 follows:
 \bqa\label{resIII}
&&\chi^2=266.9/(101-6)\ ;\nonumber\\
&&M_\kappa=898\pm 315MeV\ ,\,\,\, \Gamma_\kappa=1902\pm 282MeV\ ;\nonumber\\
&&a_0^{1/2}=0.384\pm 0.017\ ,\,\,\,a_0^{3/2}=-0.136\pm 0.002\ ;\nonumber\\
&&M_{K^*}=1468\pm 4MeV\ ,\,\,\, \Gamma_{K^*}=146\pm 23MeV\ .
 \eqa
 We see from the above equation that the $\chi_{d.o.f.}^2$ is much
 larger than the fit using only the LASS data and the scattering
 length parameter is much larger than the $\chi$PT value and the
 result should not be trustworthy.\footnote{Furthermore, for the large fit result $a_0^{1/2}$
 there is no Adler zero.} We also plot the fit results versus the $\delta_0^{1/2}$ data
 in fig.~\ref{figd1C}.
It is interesting to notice that the results of $M_\kappa$ and
$\Gamma_\kappa$ and their error bars obtained by varying
$a_0^{1/2}$, as shown in fig.~\ref{fig3a1AE}, have a very similar
behavior to the results obtained using Method I as shown in
fig.~\ref{fig3a1}, for small value of $a_0^{1/2}$. Therefore,
though not successful,  the combined fit still confirms the
conclusion that
 the $\kappa$ resonance exists if $a_0^{1/2}$ does not deviate too much from its $\chi$PT value.
\section {Summary and Conclusions}\label{conclude}
We have already made a rather long and detailed discussion  on the
new unitarization approach and the results on the $\kappa$ pole
based on the approach. In here we summarize our main physical
results: first of all, there exists the $\kappa$ resonance, if the
scattering length parameter $a_0^{1/2}$ does not deviate much from
its $\chi$PT prediction.
 Secondly, according to our fit to the LASS data, the mass of the $\kappa$ resonance is
 definitely smaller than most previous results found in the literature.
 It is found that  the width parameter  is more flexible than the mass
parameter in the fit results, which can also be seen in
fig.~\ref{fig3a1}.  Our fit results are given in Eq.~(\ref{resI}).
If we further fix $a_0^{1/2}$ to be its current $\chi$PT value
within
 1$\sigma$ error bar, the pole position of $\kappa$ is approximately given by
 Eq.~(\ref{resIa}). However, unlike $\pi\pi$ scatterings,
the scattering lengths under concern are from SU(3) $\chi$PT
results and one expects that high order corrections are more
significant here. Therefore one should also be cautious when
referring to Eq.~(\ref{resIa}) or (\ref{resIb'}). Furthermore,
there are some mismatch or disagreement between our method and
numerical results and the experimental data which predominately
come from the I=3/2 channel.\footnote{The discrepancy may be
reduced in future when a two loop $\chi$PT calculation becomes
available. This expectation is supported by a similar study on
$\pi\pi$ scatterings~\cite{z4}.} However, we have made efforts to
separate and to minimize the ambiguity when fixing the location of
the $\kappa$ pole. In fact, we are convinced by comparing
Eqs.~(\ref{resIa}) and (\ref{resIb'}) that the problem in the
I=3/2 channel has only minor influence to the $\kappa$ pole
location.

Our numerical results, especially  Eq.~(\ref{resIb'}) and the way
they are derived  are in principle consistent with that of
Ref.~\cite{CP00} in which $\kappa$ pole with similar location can
be found after adding a few `data points' generated by $\chi$PT to
the original LASS data. These `data points' should take the
similar role as constraining the scattering lengths here. The
present approach also share some similarities, at qualitative
level, with that of Ref.~\cite{Ishida97}, though in details the
two approaches differ in the treatment on both the pole
contributions and the cut contributions.

While this paper is being completed, we received a paper by
B\"uttiker, Descotes and Moussallam~\cite{Buttiker}. The central
value of $a_0^{1/2}$ given in Ref.~\cite{Buttiker} obtained by
solving Roy--Steiner equation increases considerably than that of
Ref.~\cite{Meissner}, and $a_0^{3/2}$ changes very little.
Apparent disagreement on $\delta^{3/2}_0$ between the result of
Ref.~\cite{Buttiker} and the experimental data are also indicated.
If we take roughly $a_0^{1/2}\sim 0.201 - 0.245$ as indicated by
the results of Ref.~\cite{Buttiker} (also $a_0^{3/2}$ is
readjusted to lie within the range -0.053 -- -0.038), we would
obtain, similar to obtaining Eq.~(\ref{resIb'}), table~\ref{tab3}.
\begin{table}[bt]
\centering\vspace{0.1cm}
\begin{tabular}{|c|c|c|c|c|c|}
\hline
$\Lambda^2_{L}$&$\chi^2_{tot}$&$a_0^{1/2}$&$a_0^{3/2}$&$M_\kappa$
&$\Gamma_\kappa$
\\ \hline
1.5&186.1&$0.201\pm 0.038$  &$-0.053 (\rm{at\,\, limit})$&$666\pm
11$&$565\pm 41$
\\ \hline
3&104.7 &$0.201 \pm 0.043$ &$-0.053 (\rm{at\,\, limit})$&$673\pm
10$&$562\pm 35 $
\\ \hline
5&75.2 &$0.201 \pm 0.026$ &$-0.053 (\rm{at\,\, limit})$&$681\pm
11$&$567\pm 34 $
\\ \hline
7&69.8 &$0.218\pm 0.026$ &$-0.053 (\rm{at\,\, limit})$&$682\pm
25$&$598\pm 153 $
\\ \hline
9&71.1 &$0.234 \pm 0.032$ &$-0.053 (\rm{at\,\, limit})$&$682\pm
24$&$629\pm 134 $
\\ \hline
11&74.6 &$0.243 \pm 0.032$ &$-0.053\pm 0.009$&$683\pm
16$&$651\pm 85 $
\\ \hline
20&90.1 &$0.245 \pm 0.021$ &$-0.044\pm 0.006$&$693\pm
14$&$658\pm 40 $
\\ \hline
$\infty$&137.4 &$0.245 (\rm{at\,\, limit})$ &$-0.038 (\rm{at\,\,
limit})$&$710\pm
14$&$661\pm 29 $
\\ \hline
\end{tabular}
\caption{\label{tab3}Various fit results to the LASS data up to
1.43GeV obtained by varying the cutoff parameters of the left hand
integrals and with the constraints on $a_0^{1/2}$ and $a_0^{3/2}$
from Ref.~\cite{Buttiker}. All values of $M_\kappa$ and
$\Gamma_\kappa$ in the table are in units of MeV.}
\end{table}
 Again, when there is a large $\chi^2$ occurs in table~\ref{tab3} it is mainly
  contributed by the I=3/2 channel. The `systematic' errors  induced
by varying cutoff parameters are estimated and the final results
on the $\kappa$ pole position obtained by considering the
constraints provided by Ref.~\cite{Buttiker} is estimated from
table~\ref{tab3}:
 \be
M_\kappa=688\pm 25\pm 22{\mathrm MeV}\ ,\,\,\,
\Gamma_\kappa=613\pm 153\pm 48{\mathrm MeV}\ .\ee
 Here the first
error bar corresponds to the largest error bar generated by MINUIT
when varying the cutoff parameters. The second error bar is our
estimate obtained by examining the variation of the corresponding
central value when changing the cutoff parameters, which is also a
rather conservative estimate.

{\bf\large Acknowledgement} We would like to thank Prof. Y.~S.~Zhu
for helpful discussions on data fittings. This work is supported
in part by China National Natural Science Foundation under grant
number 10047003 and 10055003.

\begin{newpage}

\end{newpage}
\end{document}